\RequirePackage{fix-cm}
\documentclass[twocolumn, epjc3, comma, sort&compress, natbib]{svjour3}
\bibpunct{[}{]}{,}{n}{}{,} % to get "[numbered]" references from natbib

\journalname{Eur. Phys. J.}

\usepackage{latexsym}
\usepackage{amsmath}
\usepackage{amssymb}
\usepackage{amsfonts}
\usepackage{CJKutf8}

\usepackage[mathscr,scaled=1.15]{urwchancal}
\DeclareFontFamily{OT1}{pzc}{}
\DeclareFontShape{OT1}{pzc}{m}{it}%
{<-> s * [1.15] pzcmi7t}{}
\DeclareMathAlphabet{\mathpzc}{OT1}{pzc}{m}{it}

\usepackage{color}

\usepackage{supertabular}
\usepackage{placeins}
\usepackage{epsfig}
\usepackage{graphicx}

\definecolor{purple}{rgb}{0.5,0,0.5}
\definecolor{blue}{rgb}{0.0,0,0.9}
\definecolor{prdblue}{rgb}{0.133,0.118,0.498}
\usepackage[colorlinks=true, pdfstartview=FitV, linkcolor=prdblue, citecolor= prdblue, urlcolor=prdblue]{hyperref}

\makeatletter

\setbox0\hbox{$\xdef\scriptratio{\strip@pt\dimexpr
    \numexpr(\sf@size*65536)/\f@size sp}$}

\newcommand{\scriptveryshortarrow}[1][3pt]{{%
    \hbox{\rule[\scriptratio\dimexpr\fontdimen22\textfont2-.2pt\relax]
               {\scriptratio\dimexpr#1\relax}{\scriptratio\dimexpr.4pt\relax}}%
   \mkern-4mu\hbox{\let\f@size\sf@size\usefont{U}{lasy}{m}{n}\symbol{41}}}}

\makeatother

\begin{document}

\begin{CJK}{UTF8}{song}

%Title of paper
%\title{Symmetry-preserving calculation of the pion's valence-quark distribution}
%\title{Unification of continuum and lattice predictions of the pion's valence distribution}
%\title{Symmetry, symmetry breaking, and pion parton distributions}
\title{$\,$\\[-6ex]\hspace*{\fill}{\normalsize{\sf\emph{Preprint nos}.\ NJU-INP 062/22, USTC-ICTS/PCFT-22-19}}\\[1ex]
%E615}
Nucleon axial form factor at large momentum transfers
}

\author{Chen Chen\thanksref{USTC1,USTC2}
        %Chen Chen (陈晨)\thanksref{USTC1,USTC2}%
        $\,^{\href{https://orcid.org/0000-0003-3619-0670}{\textcolor[rgb]{0.00,1.00,0.00}{\sf ID}}}$
%%        \and
%%        Zhu-Fang Cui (崔著钫)\thanksref{NJU,INP}%
%%        $\,^{\href{https://orcid.org/0000-0003-3890-0242}{\textcolor[rgb]{0.00,1.00,0.00}{\sf ID}}}$
       and
       Craig D.~Roberts\thanksref{NJU,INP}%
       $\,^{\href{https://orcid.org/0000-0002-2937-1361}{\textcolor[rgb]{0.00,1.00,0.00}{\sf ID}}}$
}

%\thankstext{eYL}{luya@nju.edu.cn}
%\thankstext{eDB}{binosi@ectstar.eu}
%\thankstext{eMD}{mding@ectstar.eu}
%\thankstext{eCDR}{cdroberts@nju.edu.cn}
%\thankstext{eHYX}{hyxing@smail.nju.edu.cn}
%\thankstext{eCX}{cxu@nju.edu.cn}

%\authorrunning{C.~Chen, Z.-F. Cui and C.\,D.~Roberts} % if too long for running head
\authorrunning{C.~Chen and C.\,D.~Roberts} % if too long for running head

\institute{Interdisciplinary Center for Theoretical Study, University of Science and Technology of China, Hefei, Anhui 230026, China \label{USTC1}
            \and
            Peng Huanwu Center for Fundamental Theory, Hefei, Anhui 230026, China \label{USTC2}
            \and
            School of Physics, Nanjing University, Nanjing, Jiangsu 210093, China \label{NJU}
           \and
           Institute for Nonperturbative Physics, Nanjing University, Nanjing, Jiangsu 210093, China \label{INP}
\\[1ex]
Email:
%\email[]{phycui@nju.edu.cn}
%\email[]{m.ding@hzdr.de}
%\email[]{josemanuel.morgado@dci.uhu.es}
%\email[]{khepani@ugr.es}
\href{mailto:chenchen1031@ustc.edu.cn}{chenchen1031@ustc.edu.cn} (C.~Chen);
%\href{mailto:phycui@nju.edu.cn}{phycui@nju.edu.cn} (Z.-F.~Cui);
\href{mailto:cdroberts@nju.edu.cn}{cdroberts@nju.edu.cn} (C.\,D.~Roberts).
            }

%Collaboration name if desired (requires use of superscriptaddress
%option in \documentclass). \noaffiliation is required (may also be
%used with the \author command).
%\collaboration can be followed by \email, \homepage, \thanks as well.
%\collaboration{}
%\noaffiliation

\date{2022 June 24}
%\date{2022 June 02}
%\date{2022 May 07}

\maketitle

\end{CJK}

\begin{abstract}
Using a Poincar\'e-covariant quark+diquark Faddeev equation and related symmetry-preserving \linebreak weak interaction current, we deliver parameter-free predictions for the nucleon axialvector form factor, $G_A(Q^2)$, on the domain $0\leq x=Q^2/m_N^2\leq 10$, where $m_N$ is the nucleon mass.
We also provide a detailed analysis of the flavour separation of the proton $G_A$ into contributions from valence $u$ and $d$ quarks; and with form factors available on such a large $Q^2$ domain, predictions for the flavour-separated axial-charge light-front transverse spatial density profiles.
Our calculated axial charge ratio $g_A^d/g_A^u=-0.32(2)$ is consistent with available experimental data and markedly larger in magnitude than the value typical of nonrelativistic quark models.  The value of this ratio is sensitive to the strength of axialvector diquark correlations in the Poincar\'e-covariant nucleon wave function.  Working with a realistic axialvector diquark content, the $d$ and $u$ quark transverse density profiles are similar.
Some of these predictions could potentially be tested with new data on threshold pion electroproduction from the proton at large $Q^2$.
%
%\rule{0.78\linewidth}{0.1ex}
\end{abstract}
%%
%%Keywords:
%%proton \sep
%%magnetic charge radius \sep
%%electric charge radius \sep
%%emergence of mass \sep
%%lepton-hadron scattering \sep
%%strong interactions in the standard model of particle physics

%%%%%%%%%%%%%%%%%%%%%%%%%%%%%%%%%%%%%%%%%%%%%%%%%%%%%%%%%%%%%%%%%%%%%%%%%%%%%%%%%%%%%%%%%%%%%%%%%%%%%%%%%%%%%%%%%%%%%%%
% 4500 words

%\noindent\emph{1.$\;$Introduction} ---
\section{Introduction}
Electroweak interactions of the nucleon are described by four form factors.  The Dirac and Pauli form factors, $F_{1,2}$, which are key to understanding electromagnetic interactions, have long been probed by experiment and analysed by theory \cite{Perdrisat:2006hj, Arrington:2006zm, Holt:2012gg}.  Poincar\'e-invariant predictions now exist out to momentum transfers $Q^2 \approx 20\,$GeV$^2$ \cite{Cui:2020rmu} and will be tested in forthcoming experiments \cite{Gilfoyle:2018xsa}.  Of particular interest are data on the flavour separation of these form factors \cite{Cates:2011pz, Wojtsekhowski:2020tlo}, which have the potential to validate the predicted roles of nonpointlike quark+quark (diquark) correlations within the nucleon \cite{Barabanov:2020jvn}.

The nucleon's axial current is also characterised by two form factors:
\begin{subequations}
\label{jaxdq0}
\begin{align}
\label{jaxdq}
J^j_{5\mu}(&K,Q)
:= \langle N(P_f)|{\mathpzc A}^j_{5\mu}(0)|N(P_i)\rangle \\
\label{jaxdqb}
=&\bar{u}(P_f)\frac{\tau^j}{2}\gamma_5
\bigg[ \gamma_\mu G_A(Q^2) +\frac{iQ_\mu}{2m_N}G_P(Q^2) \bigg]\,u(P_i)\,,
\end{align}
\end{subequations}
where we have assumed isospin symmetry;
$P_i$ and $P_f$ are, respectively, the initial and final momenta of the nucleon(s) involved, defined such that the on-shell condition is fulfilled, $P_{i,f}^2=-m_N^2$, with $m_N$ the nucleon mass;
$\{\tau^i|i=1,2,3\}$ are Pauli matrices;
and $K=(P_f+P_i)/2$, $Q=(P_f-P_i)$.
In Eq.\,\eqref{jaxdq0}, $G_A(Q^2)$ is the axial form factor and $G_P(Q^2)$ is the induced pseudoscalar form factor.  These two functions are far less well known than $F_{1,2}$.

This is an issue because, amongst other things, the analysis and reliable interpretation of modern neutrino experiments relies on sound theoretical knowledge of neutrino/antineutrino-nucleus ($\nu/\bar \nu$-$A$)
interactions \cite{Mosel:2016cwa, Alvarez-Ruso:2017oui, Hill:2017wgb, Gysbers:2019uyb, Lovato:2020kba}; and a crucial element in such calculations is the nucleon axial form factor, $G_A(Q^2)$.  The $Q^2=0$ value of $G_A$ is the nucleon's isovector axial charge, $g_A=1.2756(13)$ \cite{Zyla:2020}, which determines the rate of neutron-to-proton $\beta$-decay: $n\to p + e^- + \bar\nu$.  At the structural level, $g_A$ measures the difference between the light-front number-densities of quarks with helicity parallel and antiparallel to that of the nucleon \cite{Deur:2018roz}.

Regarding $G_P(Q^2)$, calculations show that the pion pole dominance \emph{Ansatz}:
\begin{equation}
\label{ppd}
G_P \simeq \frac{4m_N^2G_A}{Q^2+m_\pi^2} \,,
\end{equation}
where $m_\pi$ is the pion mass, is an excellent approximation \cite{Jang:2019vkm, Bali:2019yiy, Chen:2020wuq, Chen:2021guo}.  Hence, one may focus on the axial form factor, $G_A(Q^2)$.

$G_A(Q^2)$ has long attracted interest.  It was extracted from $\nu p$ and $\nu$-deuteron, $d$, scattering experiments performed over thirty years ago \cite{Baker:1981su, Miller:1982qi, Kitagaki:1983px, Ahrens:1986xe}, with results that are consistent with dipole behaviour characterised by a mass-scale $M_A \approx 1.1\,m_N$.  However, these cross-sections only extend to $Q^2\lesssim 1\,$GeV$^2$.  A new analysis of the world's data on $\nu d$ scattering, which reaches to $Q^2\lesssim 3\,$GeV$^2$, returns a similar value \cite{Meyer:2016oeg}, albeit with larger uncertainty.
In contrast, experiments using $\nu$ scattering on an array of heavier targets (water, iron, mineral oil, Kevlar, and carbon) yield results covering the range $1.1 \lesssim M_A/m_N \lesssim 1.60$ \cite{Gran:2006jn, Dorman:2009zz, AguilarArevalo:2010zc, Fields:2013zhk, Fiorentini:2013ezn}.  Regarding these extractions, issues relating to the reliability of the model employed to describe the nuclear target and differences between the models used by the collaborations must be considered.

Theoretically, there is no reason why $G_A(Q^2)$ should be well-approximated by a single dipole on the entire domain of spacelike $Q^2$; and deviations from this behaviour may become important as experimental facilities begin to probe weak interactions at larger momentum transfers.
For instance, precise knowledge of the larger $Q^2$ behaviour of $G_A(Q^2)$ is important for experimental extraction of the proton's strange-quark form factor \cite{Wojtsekhowski:2020tlo}.

\begin{figure}[t]
\centerline{%
\includegraphics[clip, height=0.14\textwidth, width=0.45\textwidth]{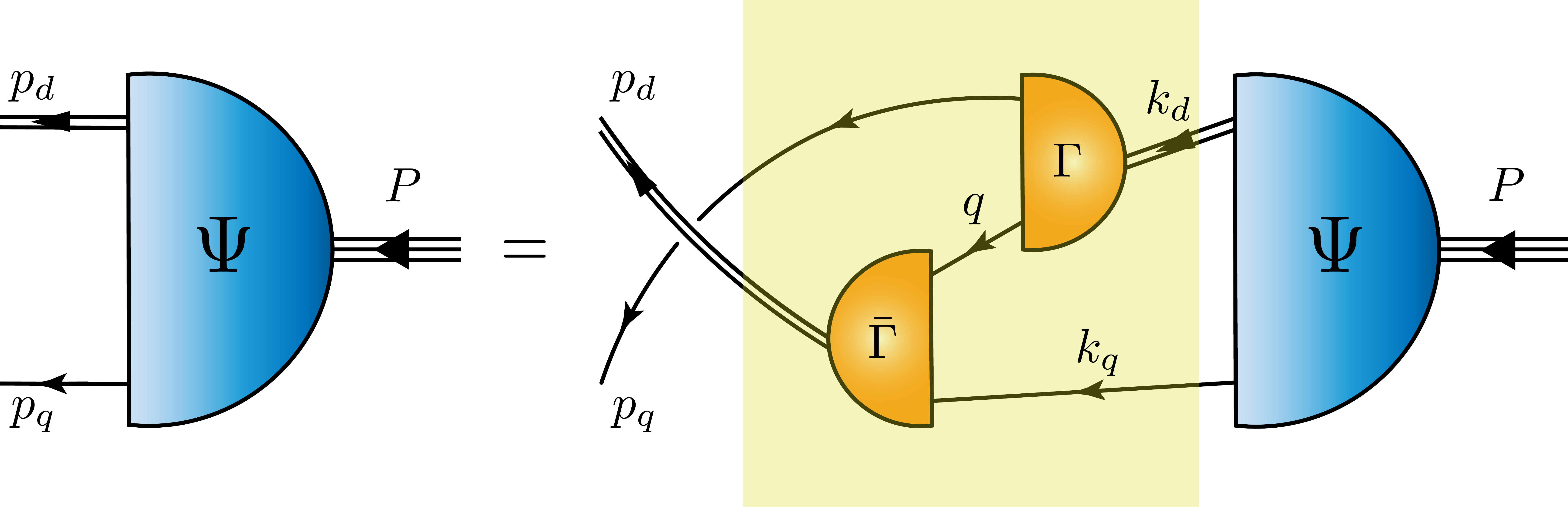}}
\caption{\label{figFaddeev}
Linear integral equation for the Poincar\'e-covariant matrix-valued function $\Psi$, the Faddeev amplitude for a nucleon with total momentum $P=p_q+p_d=k_q+k_d$ constituted from three valence quarks, two of which are always contained in a nonpointlike diquark correlation. $\Psi$ describes the relative momentum correlation between the dressed-quarks and -diquarks. Legend. \emph{Shaded rectangle} -- Faddeev kernel; \emph{single line} -- dressed-quark propagator; $\Gamma$ -- diquark correlation amplitude; and \emph{double line} -- diquark propagator. Ground-state nucleons ($n$ - neutron, $p$ - proton) contain both isoscalar-scalar diquarks, $[ud]\in(n,p)$, and isovector-axialvector diquarks $\{dd\}\in n$, $\{ud\}\in (n,p)$, $\{uu\}\in p$.}
\end{figure}

Furthermore, it is conceivable that the study of \linebreak threshold pion electroproduction at large momentum transfers could yield data on $G_A(Q^2)$ that reaches out to $Q^2\lesssim 10\,$GeV$^2$ \cite{Anikin:2016teg}; and, in fact, data have already been obtained on $2\lesssim Q^2/{\rm GeV}^2 \lesssim 4$ \cite{CLAS:2012ich}.  Were this promise to be fulfilled, then one would be in a position to test the dipole \emph{Ansatz} and, importantly, deliver empirical information on the distribution and flavour-separation of axial charge within the nucleon.

A computation of the large-$Q^2$ behaviour of $G_A(Q^2)$ was completed using light-cone sum rules in Ref.\,\cite{Anikin:2016teg}.  It closed with the suggestion that the results obtained could be confronted with predictions from quantum chromodynamics (QCD) obtained using lattice regularisation or continuum Schwinger function methods (CSMs) \cite{Eichmann:2016yit, Qin:2020rad}.  At that time, the large-$Q^2$ domain was a challenge to both approaches.  Whilst this remains the case for lattice-QCD (lQCD), recent advances in computer hardware and progress with formulating the nucleon axial-current problem \cite{Chen:2020wuq, Chen:2021guo} mean that a CSM calculation of $G_A(Q^2)$ is now possible on a large domain of spacelike $Q^2$; and potentially, with continuing algorithm improvement, the entire domain.

\begin{figure}[!t]
\centerline{\includegraphics[clip, width=0.49\textwidth]{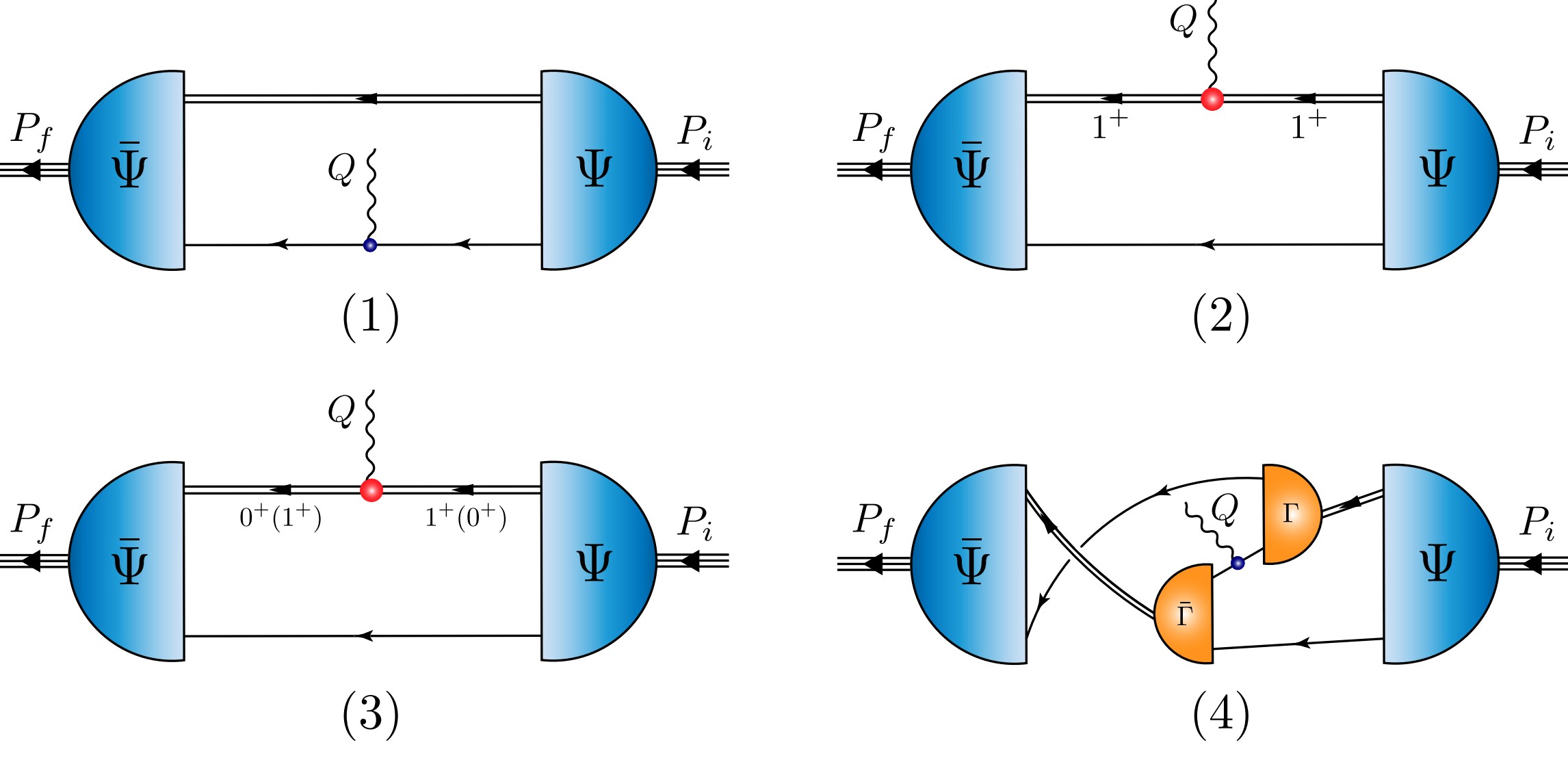}}
\caption{\label{figcurrent} Symmetry-preserving axial current for on-shell baryons described by the Faddeev amplitudes produced by the equation depicted in Fig.\,\ref{figFaddeev}: \emph{single line}, dressed-quark propagator; \emph{undulating line}, the axial or pseudoscalar current; $\Gamma$,  diquark correlation amplitude; \emph{double line}, diquark propagator.  For $G_A(Q^2)$, seagull diagrams do not contribute \cite{Chen:2021guo}.}
\end{figure}

Section~\ref{SecFE} sketches the nucleon Faddeev equation and associated symmetry-preserving axial current that form the basis for the analysis herein.  Our predictions for the nucleon axial form factor are described in Sec.\,\ref{SecResults} as are their comparisons with data and comparable computations.  A detailed discussion of the flavour-sepa-rated proton axial form factors is presented in Sec.~\ref{SecFlavourSep}, including results for the associated light-front transverse spatial density profiles.  Section~\ref{epilogue} provides a summary and perspective.

\section{Faddeev equation and axial current}
\label{SecFE}
Our calculation of $G_A(Q^2)$ rests on a solution of the Poincar\'e-covariant Faddeev equation depicted in Fig.\,\ref{figFaddeev}, which, when inserted into the diagrams drawn in Fig.\,\ref{figcurrent}, delivers a result for the current in Eq.\,\eqref{jaxdq0} that ensures, \emph{inter alia}, partial conservation of the axialvector current (PCAC) and the associated Goldberger-Treiman relation; hence, realistic predictions for $G_A(Q^2)$.  Details are presented in Refs.\,\cite{Chen:2020wuq, Chen:2021guo}, which provided results for $G_A(Q^2)$ on $Q^2\in [0,1.8]\,$GeV$^2$ and numerous comparisons with existing data and modern calculations.  Such calculations are largely restricted to low-$Q^2$.
For subsequent use, we identify the following separations of the current in Fig.\,\ref{figcurrent}.
\begin{enumerate}
\item Diagram (1), two distinct terms: $\langle J \rangle^{S}_{\rm q}$ -- weak-boson strikes dressed-quark with scalar diquark spectator; and $\langle J \rangle^{A}_{\rm q}$ -- weak-boson strikes dressed-quark with axialvector diquark spectator.
\item Diagram (2): $\langle J \rangle^{AA}_{\rm qq}$ -- weak-boson strikes axialvector diquark with dressed-quark spectator.
\item Diagram (3): $\langle J \rangle^{\{SA\}}_{\rm qq}$ -- weak-boson mediates transition between scalar and axialvector diquarks, with dressed-quark spectator.
\item Diagram (4), three terms:
    $\langle J \rangle_{\rm ex}^{SS}$ -- weak-boson strikes dressed-quark ``in-flight'' between one scalar diquark correlation and another;
    $\langle J \rangle_{\rm ex}^{\{SA\}}$ -- dressed-quark ``in-flight'' between a scalar diquark correlation and an axialvector correlation;
    and $\langle J \rangle_{\rm ex}^{AA}$ -- ``in-flight'' between one axialvector correlation and another.
\end{enumerate}

Supplying a little more background, then regarding Fig.\,\ref{figFaddeev} and accounting for Fermi-Dirac statistics, five types of dynamical diquark correlations are possible in a $J=1/2$ bound-state.  However, only isoscalar-scalar, isovector-axial\-vec\-tor are quantitatively important in ground-state positive-parity systems \cite{Barabanov:2020jvn, Yin:2021uom, Raya:2021pyr, Eichmann:2022zxn}.  We use the following values for their masses:
{\allowdisplaybreaks
\begin{subequations}
\label{dqmasses}
\begin{align}
m_{[ud]_{0^+}} &= 0.80\,\mbox{GeV}\,,\\
m_{\{uu\}_{1^+}} = m_{\{ud\}_{1^+}} = m_{\{dd\}_{1^+}} & = 0.89\,\mbox{GeV}\,,
\end{align}
\end{subequations}
drawn from Ref.\,\cite{Segovia:2014aza}.  The dressed light-quarks are characterised by a Euclidean constituent mass $M_q^E = 0.33$ GeV.  All associated propagators and additional details concerning the Faddeev kernel are presented in Ref.\,\cite[Appendix~A]{Chen:2021guo}.  With these inputs, one obtains a good description of many dynamical properties of baryons \cite{Burkert:2017djo, Chen:2018nsg, Lu:2019bjs, Cui:2020rmu, Lu:2022cjx}.
%% the nucleon, $\Delta$-baryon and Roper resonance.  %Herein, we report a model uncertainty obtained by independently varying the diquark masses by $\pm 5$\%, thereby changing $m_N$ by $\pm 3$\%.
}

The solution of the Faddeev equation yields the nucleon Faddeev amplitude, $\Psi$, and mass, $m_N = 1.18\,$GeV.  This value is intentionally large because Fig.\,\ref{figFaddeev} describes the nucleon's \emph{dressed-quark core}.  To produce the complete nucleon, resonant contributions should be included in the Faddeev kernel.  Such ``meson cloud'' effects generate the physical nucleon, whose mass is roughly $0.2$\,GeV lower than that of the core \cite{Hecht:2002ej, Sanchis-Alepuz:2014wea}.  (Similar effects are reported in quark models \cite{Garcia-Tecocoatzi:2016rcj, Chen:2017mug}.)  Their impact on nucleon structure can be incorporated using dynamical coupled-channels models \cite{Aznauryan:2012ba, Burkert:2017djo}, but that is beyond the scope of contemporary Faddeev equation analyses.
Instead, we express all form factors in terms of $x= Q^2/m_N^2$, a procedure that has proved efficacious in developing solid comparisons with experiment \cite{Burkert:2017djo, Chen:2018nsg, Lu:2019bjs, Cui:2020rmu}.
In addition, we report an uncertainty obtained by independently varying the diquark masses by $\pm 5$\%, thereby changing $m_N$ by $\pm 3$\%.
%, and combining the obtained results with weight determined by the relative strength of scalar and pseudovector diquark contributions to $G_A(0)$, \emph{i.e}.\ 4:1.  Notably, scalar and pseudovector diquark variations interfere destructively, \emph{e.g}.\ reducing $m_{[ud]}$ increases $G_A(0)$, whereas the same change in the pseudovector mass decreases it.

\begin{figure}[t]
\vspace*{2ex}

\leftline{\hspace*{0.5em}{\large{\textsf{A}}}}
\vspace*{-4ex}
\includegraphics[width=0.42\textwidth]{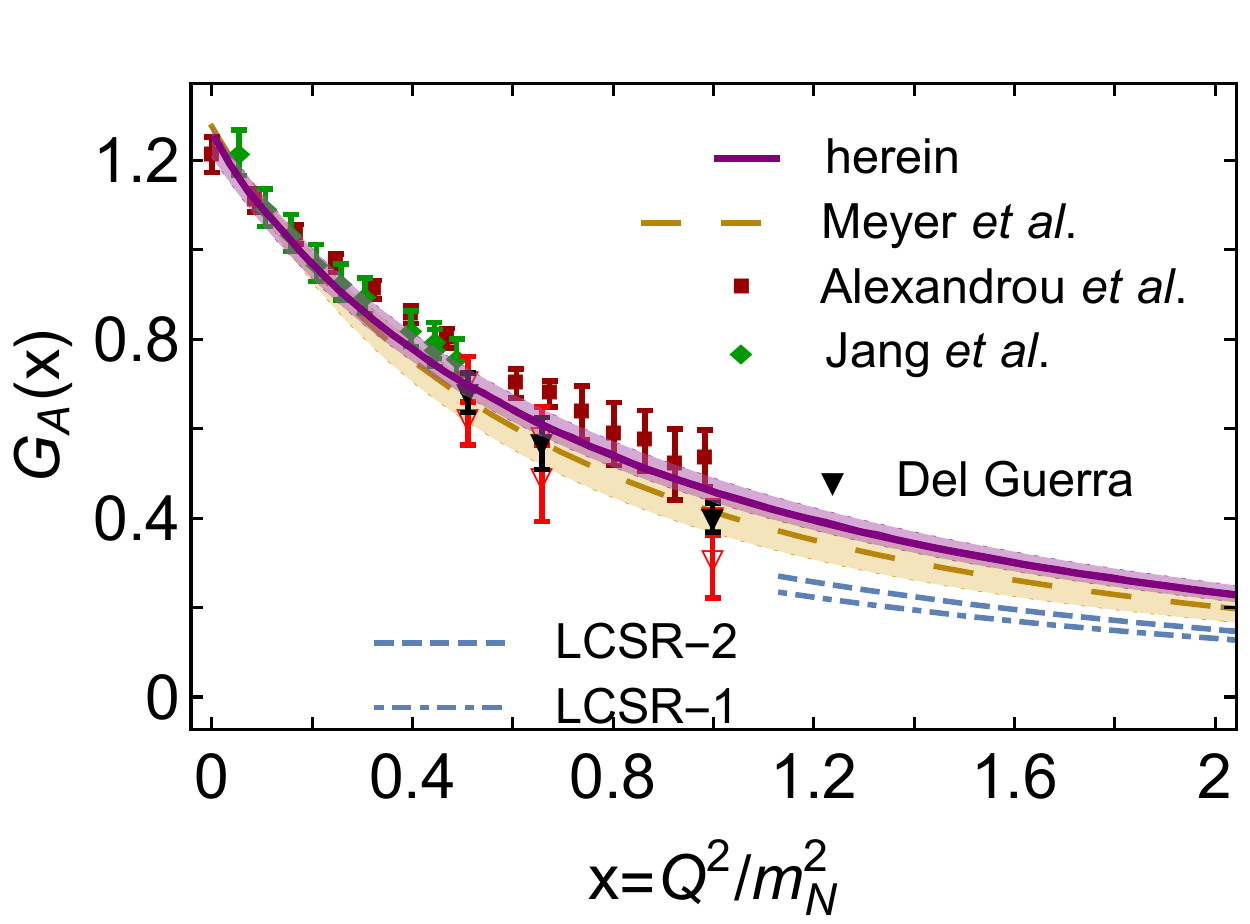}
\vspace*{0ex}

\leftline{\hspace*{0.5em}{\large{\textsf{B}}}}
\vspace*{-4ex}
\includegraphics[width=0.42\textwidth]{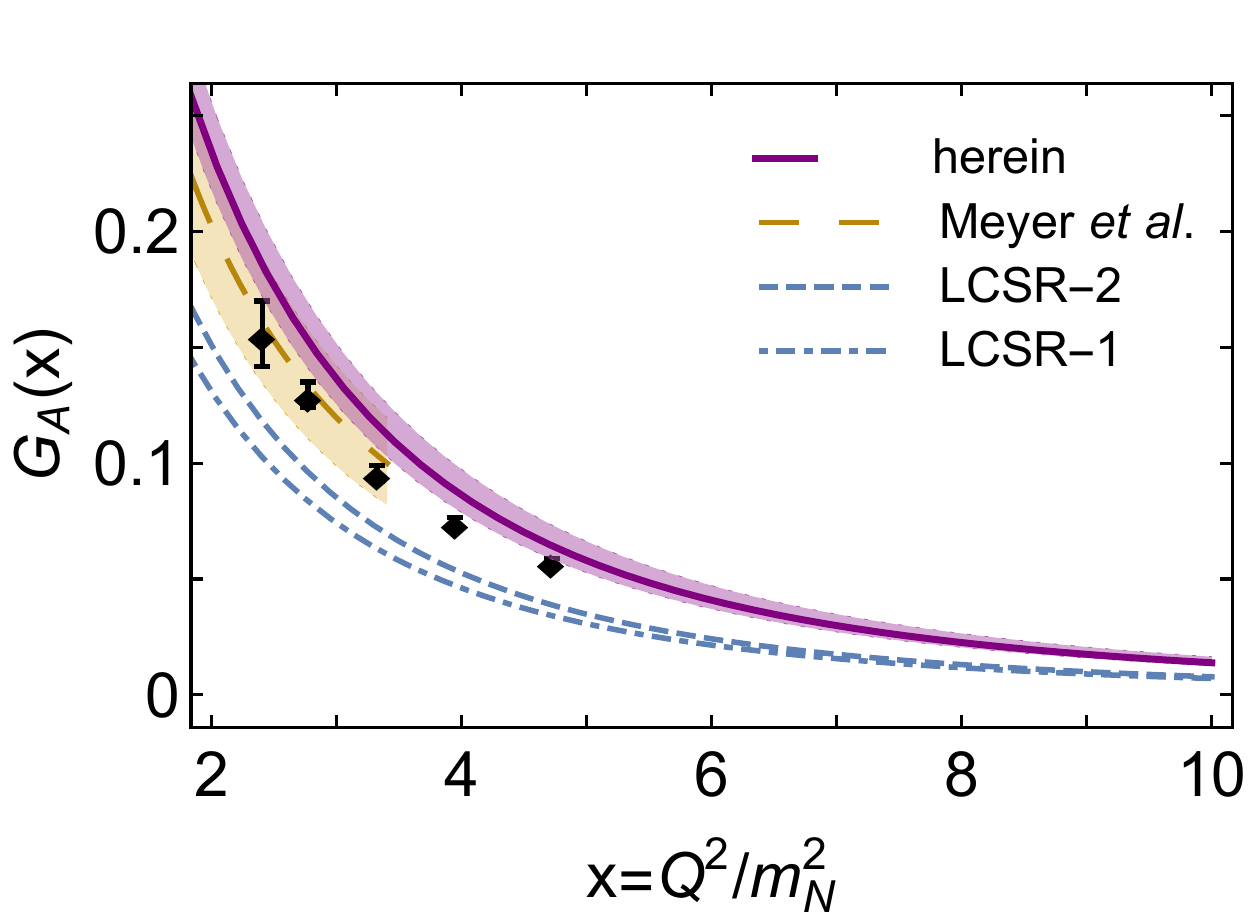}
\caption{\label{FigGAx}
{\sf Panel A}.
Low-$x$ behaviour of $G_A(x=Q^2/m_N^2)$.
Solid purple curve -- CSM prediction.  The associated shaded band shows the sensitivity to $\pm 5$\% variation in diquark masses, Eq.\,\eqref{dqmasses}.
Comparisons provided with:
dipole fit to data -- long-dashed gold curve within like-coloured band \cite{Meyer:2016oeg};
values deduced from threshold pion electroproduction data under three different assumptions \cite{DelGuerra:1976uj} -- open and filled triangles;
results from lattice QCD --
green diamonds \cite{Jang:2019vkm} and
red squares \cite{Alexandrou:2017hac};
and light-cone sum rule results \cite{Anikin:2016teg} obtained using two models of the nucleon distribution amplitudes \cite[Table~I]{Anikin:2013aka}. -- LCSR-1 (dot-dashed blue curve) and LCSR-2 (dashed blue curve).  The LCSR approach cannot deliver a result below $x \simeq 1$.
{\sf Panel B}.
Large-$x$ behaviour of $G_A(x)$.
Curves drawn according to the Panel A legend with the addition of threshold pion electroproduction data from Ref.\,\cite{CLAS:2012ich} -- black diamonds.
}
\end{figure}

\section{Nucleon axial form factor}
\label{SecResults}
Our prediction for $G_A(x)$ is displayed in Fig.\,\ref{FigGAx}.  On $0\leq x\leq 10$, the central result is reliably interpolated using
\begin{equation}
\label{GAcentral}
%G_A(x) = \frac{1.248 + 0.060 x}{1+ 1.447 x + 0.312 x^2 + 0.088 x^3}\,.
G_A(x) = \frac{1.248 + 0.039 \,x}{1+ 1.417 \,x + 0.318 \,x^2 + 0.071 \,x^3}\,.
\end{equation}
%% \frac{0.0603825 x+1.24819}{0.0884578 x^3+0.31215 x^2+1.44747 x+1}
%% \frac{0.0394482 x+1.24829}{0.0713238 x^3+0.318342 x^2+1.41746 x+1}

First consider the low-$x$ behaviour -- Fig.\,\ref{FigGAx}A, wherein the dipole fit to data from Ref.\,\cite{Meyer:2016oeg} is also drawn: $m_A = 1.15(8) m_N$.  The dipole's domain of validity is constrained to $x\lesssim 3$; on this domain, it is a fair match for the CSM prediction.  This can be quantified by noting that were we to fit our result on $x\leq 3$ using a dipole, then the procedure would yield $m_A = 1.24(3) m_N$.
Lat\-tice-QCD values are available on $0\leq x \leq 1$ \cite{Alexandrou:2017hac, Jang:2019vkm} whereupon they agree with the CSM result.  For instance, measuring the Ref.\,\cite{Jang:2019vkm} points against our central result, the mean-$\chi^2$ is $0.79$.

The large-$x$ behaviour of $G_A(x)$ is highlighted in Fig.\,\ref{FigGAx}B.  Comparing our prediction with available data on this domain \cite{CLAS:2012ich}, there is agreement within mutual uncertainties.  It is worth stressing that our results are parameter-free predictions; especially because the framework employed is precisely the same as that used elsewhere in the explanation and prediction of
nucleon elastic form factors \cite{Cui:2020rmu}
and the electroexcitation amplitudes of the $\Delta(1232)\tfrac{3}{2}^+$ and $N(1440)\tfrac{1}{2}^+$ \cite{Carman:2020qmb, Burkert:2019bhp, Mokeev:2022xfo}.  Hence, the agreement is meaningful, providing support for the picture of emergent hadron mass expressed in our formulation of the baryon bound-state problem \cite{Roberts:2020hiw, Binosi:2022djx, Aguilar:2021uwa}.
Future experiments using modern detectors, like SuperBigBite (SBS) and CLAS12 at Jefferson Lab, can push knowledge of elastic form factors and electroexcitation amplitudes beyond $Q^2 = 5\,$GeV$^2$, a domain on which CSM predictions are already available \cite{Lu:2019bjs}.
Thus, the results in Fig.\,\ref{FigGAx}B provide strong motivation for complementing studies of ground- and excited-state nucleon structure using the vector electromagnetic current by the determination of $G_A(Q^2)$ at large $Q^2$ using near-threshold pion electroproduction from the proton, which will expose nucleon structure as seen by a hard axial current probe.

\begin{figure}[t]
\vspace*{2ex}

\leftline{\hspace*{0.5em}{\large{\textsf{A}}}}
\vspace*{-4ex}
\includegraphics[width=0.42\textwidth]{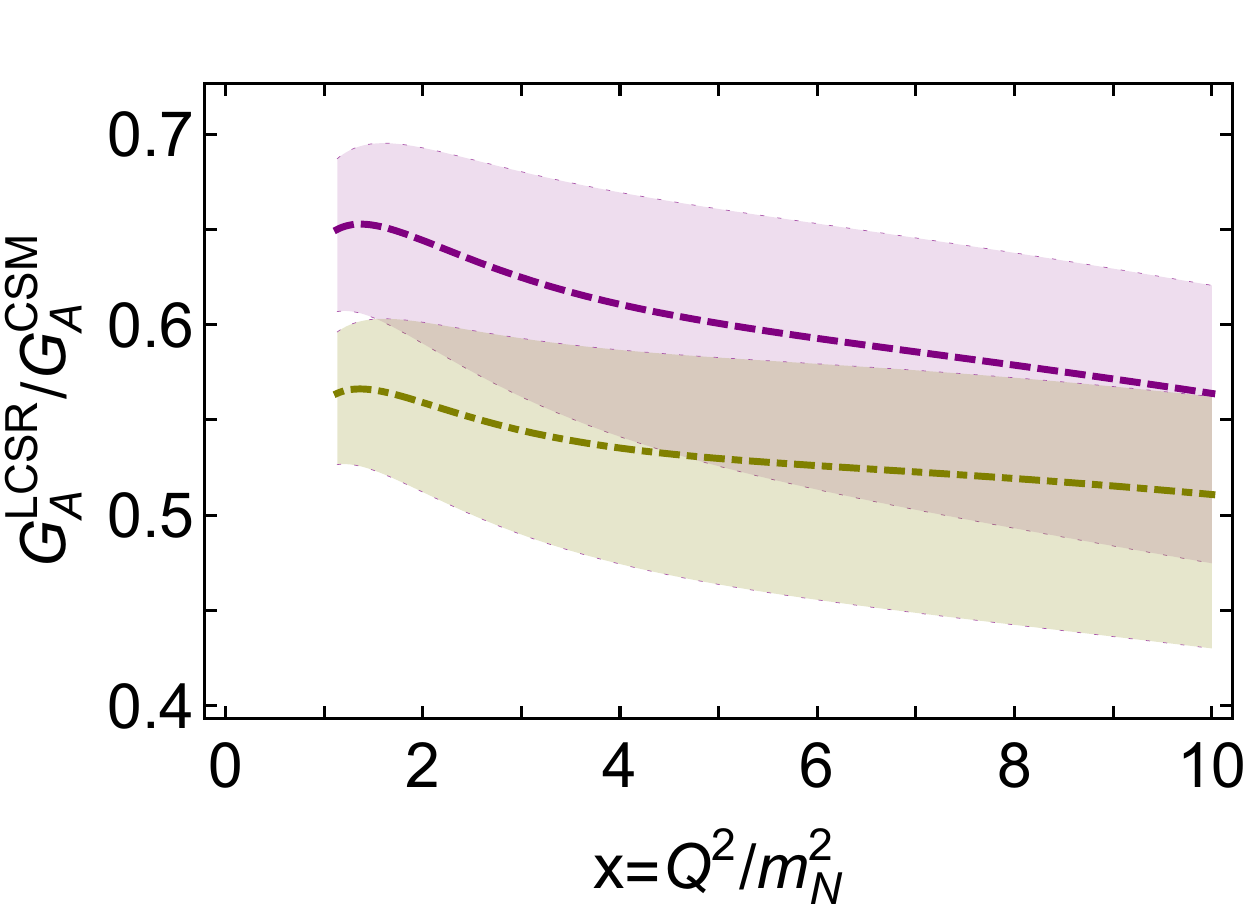}
\vspace*{0ex}

\leftline{\hspace*{0.5em}{\large{\textsf{B}}}}
\vspace*{-4ex}
\includegraphics[width=0.42\textwidth]{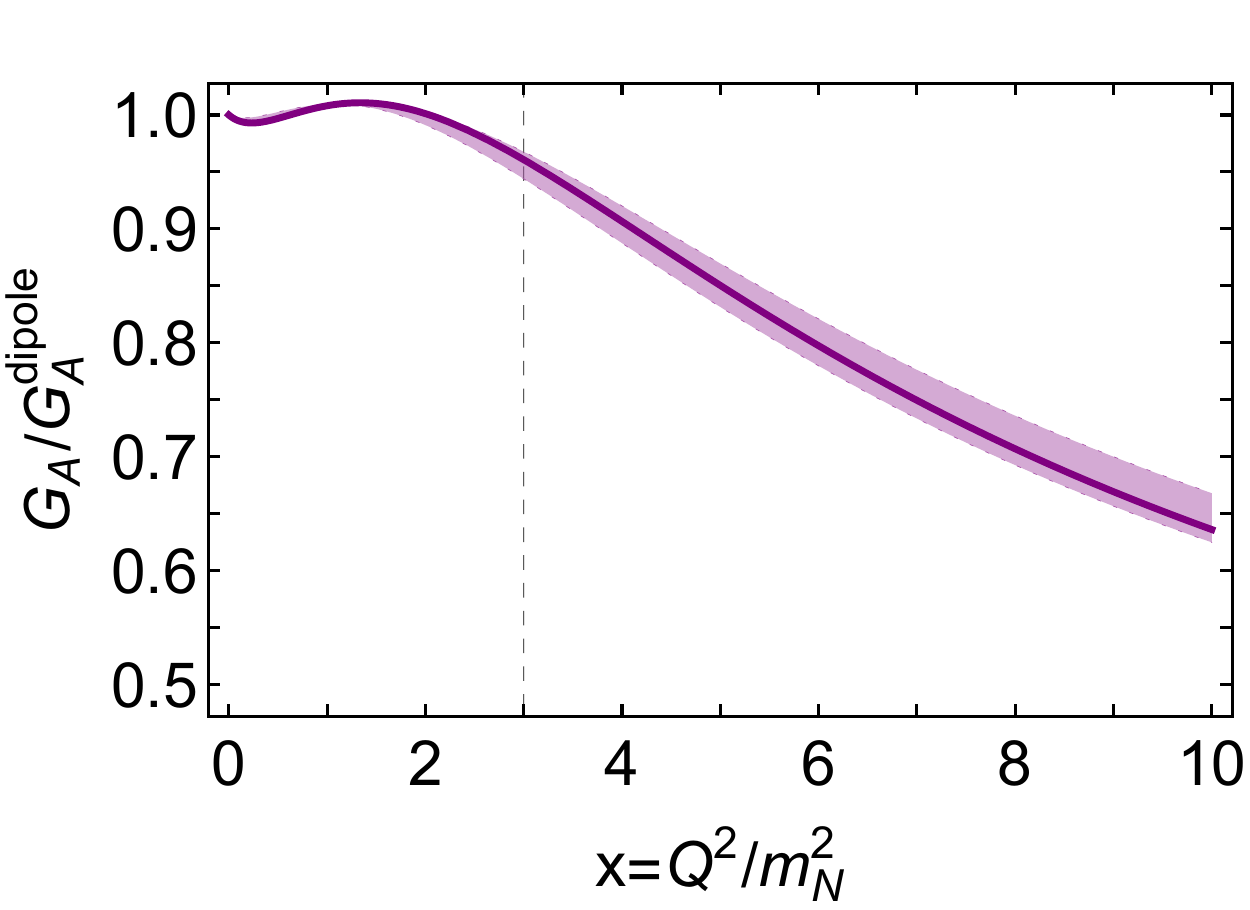}
\caption{\label{LCSRdiff}
{\sf Panel A}.
LCSR results for $G_A(x)$ compared with the CSM predictions, as measured by ratios of the associated curves in Fig.\,\ref{FigGAx}B.  The mean value of the ratio is $0.56(7)$.
{\sf Panel B}.
CSM prediction for $G_A(x)$ compared with a dipole fitted to the result on $x\in[0,3]$.  Whilst the dipole is reasonable within the fitted domain, its reliability as an approximation rapidly deteriorates as $x$ is increased beyond $x=3$.  At $x=10$, the value of the drawn ratio is $0.64(2)$.
}
\end{figure}

Fig.\,\ref{FigGAx}B also displays the only other existing $G_A$ calculation \cite{Anikin:2016teg} that extends to $x\approx 10$.  It used light cone sum rules (LCSRs); hence, cannot deliver form factor values below $x\approx 1$.  As highlighted in Fig.\,\ref{LCSRdiff}A, the LCSR results for $G_A(x)$ are markedly different from our prediction.  The mismatch is roughly a uniform factor of two, with the LCSR result lying below.

%Light-cone sum rule results are shown for two models of the leading- and higher-twist nucleon DAs, ABO1 (solid curves) and ABO2 (dashed curves), specified in Ref.\,\cite[Table~I]{Anikin:2013aka}.

%deduced from threshold pion electroproduction data under different assumptions \cite{DelGuerra:1976uj}

It is also worth comparing our prediction for $G_A(x)$ with an extrapolation of a typical dipole fit, which might be considered a reasonable tool for developing cross-section estimates in the absence of a viable alternative.  As noted above, when a dipole is used to fit our results on $x\leq 3$, one obtains a dipole mass $m_A = 1.24(3) m_N$.  The comparison is shown in Fig.\,\ref{LCSRdiff}B.  On the fitting domain, the mean absolute relative difference between Eq.\,\eqref{GAcentral} and dipole is $1.0(0.9)$\%.  Plainly, however, the dipole increasingly overestimates the actual result as $x$ increases, being $56(5)$\% too large at $x=10$.

\section{Flavour-separated axial form factor}
\label{SecFlavourSep}
It is a longstanding prediction of Faddeev equation studies that the nucleon contains both isoscalar-scalar and isovector-axialvector diquarks correlations \cite{Barabanov:2020jvn}.  Experiment \cite[MARATHON]{Abrams:2021xum} has confirmed the predicted ratio of their relative strengths \cite{Cui:2021gzg}; showing, furthermore, that the probability that a scalar-diquark-only model of the proton might be compatible with the data is $\approx 1/7\,000\,000$.  Nevertheless, such models are still in widespread use owing largely to their simplicity.

The proton's axial form factor can be written
\begin{equation}
G_A(Q^2) = G_A^u(Q^2) - G_A^d(Q^2)\,,
\end{equation}
where $G_A^f$, $f=u,d$, are the contributions from each valence-quark flavour to the total form factor.  They may be determined by
focusing on the neutral current ($j=3$) detailed in Ref.\,\cite{Chen:2021guo},
making use of the expedient $\tau^3/2 \to {\cal Q} = {\rm diag}[{\cal Q}_u,{\cal Q}_d]$,
then tracking the flow of flavour in Ref.\,\cite[Eqs.\,(C4)\,-\,(C6)]{Chen:2021guo}.
(For weak interactions, ${\cal Q}_d/{\cal Q}_u=-1$.  The scheme has also been exploited in producing flavour separations of electromagnetic form factors by choosing appropriate charge values \cite{Wilson:2011aa, Segovia:2016zyc, Chen:2018nsg, Cui:2020rmu}.)

This analysis reveals the following diagram contributions to the individual $u$, $d$ axial form factors:
{\allowdisplaybreaks\begin{subequations}
\label{FlavourSep}
\begin{align}
G_A^u & = \langle J \rangle^{S}_{\rm q} - \langle J \rangle^{A}_{\rm q} + \langle J \rangle^{AA}_{\rm qq} + \tfrac{1}{2} \langle J \rangle^{\{SA\}}_{\rm qq} \nonumber \\
& \quad  + 2 \langle J \rangle^{\{SA\}}_{\rm ex} + \tfrac{4}{5} \langle J \rangle^{AA}_{\rm ex},\\
-G_A^d & = 2 \langle J \rangle^{A}_{\rm q} + \tfrac{1}{2}\langle J \rangle^{\{SA\}}_{\rm qq} \nonumber \\
& \quad + \langle J \rangle^{SS}_{\rm ex} - \langle J \rangle^{\{SA\}}_{\rm ex} + \tfrac{1}{5} \langle J \rangle^{AA}_{\rm ex},
\end{align}
\end{subequations}
where we have used the nomenclature introduced at the beginning of Sec.\,\ref{SecFE}.
(These expressions correct those used in Ref.\,\cite{Chen:2020wuq}, wherein sign errors were made in separating some terms.)
%% The Holt-Schmidt results are for DFs at x=1, not the moments!
%
Identified according to Eqs.\,\eqref{FlavourSep}, the calculated $Q^2=0$ contributions are listed in Table~\ref{isovectorcharge}.}

\begin{table}[t]
\caption{\label{isovectorcharge}
Diagram and flavour separation of the proton axial charge: $g_A^u=G_A^u(0)$, $g_A^d=G_A^d(0)$; $g_A^u - g_A^d = 1.25(3)$.
The listed uncertainties in the tabulated results reflect the impact of $\pm 5$\% variations in the diquark masses in Eq.\,\eqref{dqmasses}, \emph{e.g}.\ $0.88_{6_\mp} \Rightarrow 0.88 \mp 0.06$. }
\begin{center}
\begin{tabular*}%{|c|c|c|c|c|c|c|}\hline
{\hsize}
{
r@{\extracolsep{0ptplus1fil}}
|l@{\extracolsep{0ptplus1fil}}
l@{\extracolsep{0ptplus1fil}}
l@{\extracolsep{0ptplus1fil}}
l@{\extracolsep{0ptplus1fil}}
l@{\extracolsep{0ptplus1fil}}
l@{\extracolsep{0ptplus1fil}}
l@{\extracolsep{0ptplus1fil}}}\hline
 & $\langle J \rangle^{S}_{\rm q}$  & $\langle J \rangle^{A}_{\rm q}$ &$\langle J \rangle^{AA}_{\rm qq}$ & $\langle J \rangle^{\{SA\}}_{\rm qq}$
 & $\langle J \rangle_{\rm ex}^{SS}$
 & $\langle J \rangle_{\rm ex}^{\{SA\}}$
 & $\langle J \rangle_{\rm ex}^{AA}$  \\\hline
 $g_A^u$ & $0.88_{6_\mp}$ & $-0.08_{0_\pm} $ & $0.03_{0_\pm}$ & $0.08_{0_\mp}$ & $0$ & $\approx 0$ & $0.03_{\pm 1}$ \\
 $-g_A^d$ & $0$ & $\phantom{-}0.16_{0_\pm} $ & $0$ & $0.08_{0_\mp}$ & $0.05_{1_\pm}$ & $\approx 0$ & $0.01_{\pm 0}$\\  \hline
\end{tabular*}
\end{center}
\end{table}

Importantly, Eqs.\,\eqref{FlavourSep} express the fact that since a scalar diquark cannot couple to an axialvector current, Diagram\,1 in Fig.\,\ref{figcurrent} only generates a $u$-quark contribution to the proton $G_A(Q^2)$, \emph{viz}.\ $\langle J \rangle^{S}_{\rm q}$.  Hence, in a scalar-diquark-only proton, a $d$-quark contribution can only arise from Fig.\,\ref{figcurrent}\,-\,Diagram\,4, \emph{i.e}., $\langle J \rangle^{SS}_{\rm ex}$; and \linebreak $|\langle J \rangle^{SS}_{\rm ex}/\langle J \rangle^{S}_{\rm q}| \approx 0.06$.  It is also noteworthy that many scalar-diquark-only models omit Diagram~4, in which cases there is no $d$-quark contribution to $G_A(Q^2)$.

Working with the $Q^2=0$ values of the flavour-separated contributions to the proton isovector axial charge, one has
\begin{equation}
\label{gAhelicity}
g_A = g_A^u-g_A^d = \int_0^1 dx\, [\Delta u(x) - \Delta d(x)]\,,
\end{equation}
where $\Delta f(x)$ is the $f$ valence-quark's contribution to the proton's light-front helicity, \emph{viz}.\ the difference between the light-front number-density of $f$-quarks with helicity parallel to that of the proton and the kindred density with helicity antiparallel.
%The difference in Eq.\,\eqref{gAhelicity} is renormalisation group invariant, but the individual flavour-separated terms are not.
Using the solution of the Faddeev equation in Fig.\,\ref{figFaddeev}, one finds
\begin{subequations}
\label{Separations}
\begin{align}
g_A^u/g_A & = \phantom{-} 0.76 \pm 0.01\,,\\
g_A^d/g_A & =- 0.24 \pm 0.01 \,,\\
g_A^d/g_A^u & = - 0.32 \pm 0.02 \,. \label{SeparationsC}
\end{align}
\end{subequations}
It is here worth recalling a textbook result, \emph{viz}.\ $g_A^d/g_A^u = -1/4$ in nonrelativistic quark models with uncorrelated wave functions; hence, the highly-correlated proton wave function obtained as a solution of the Faddeev equation in Fig.\,\ref{figFaddeev} lodges a significantly larger fraction of the proton's light-front helicity with the valence $d$ quark.  The discussion following Eqs.\,\eqref{FlavourSep} reveals that this outcome owes to the presence of axialvector diquarks in the proton.  Namely, the fact that the current contribution arising from the $\{uu\}$ correlation, in which the probe strikes the valence $d$ quark, is twice as strong as that from the $\{ud\}$ correlation, in which it strikes the valence $u$ quark.  The relative negative sign means this increases $|g_A^d|$ at a cost to $g_A^u$.

Regarding experiment, if one assumes SU$(3)$-flavour symmetry in analyses of the axial charges of octet bary-ons, then these charges are expressed in terms of two low-energy constants,  $D$, $F$; and $g_A^u = 2 F$, $g_A^d=F-D$.  Using contemporary empirical information \cite{Zyla:2020zbs}, one finds $D= 0.774(26)$, $F=0.503(27)$ and obtains the following estimates: $g_A^u/g_A = 0.79(4)$, $g_A^d/g_A= -0.21(3)$, $g_A^d/g_A^u= -0.27(4)$, which match the predictions in \linebreak Eqs.\,\eqref{Separations} within mutual uncertainties.

Recent lQCD analyses also report results for the ratio of flavour-separated charges:
$g_A^d/g_A^u = -0.40(2)$ \cite{Bhattacharya:2016zcn};
$g_A^d/g_A^u = -0.58(3)$ \cite{Alexandrou:2019brg}.
These values were computed at a resolving scale $\zeta = 2\,$GeV.
Whilst the difference in Eq.\,\eqref{gAhelicity} is renormalisation group invariant, the individual flavour-separated terms are not; and the magnitude of $g_A^d/g_A^u$ increases under QCD evolution \cite{Dokshitzer:1977sg, Gribov:1972, Lipatov:1974qm, Altarelli:1977}.
However, the evolution of $g_A^d$, $g_A^u$ is slow because it is modulated by the size of SU$(3)$-flavour symmetry breaking \cite{Deur:2018roz}.
Consequently, even though our predictions are properly associated with the hadronic scale, $\zeta_H \approx 0.33\,$GeV \cite{Cui:2020dlm, Cui:2020tdf}, and evolution is required for unambiguous comparison with the lQCD results, there does appear to be tension between the lQCD results on one hand and, on the other, our predictions and the empirical inferences: the lQCD values seem too large in magnitude.

The ratio $g_A^d/g_A^u$ is also interesting for another reason.  In nonrelativistic quark models, the helicity and transversity distributions are identical because boosts and rotations commute with the Hamiltonian; hence, $g_A^d/g_A^u = g_T^d/g_T^u$, where $g_T^{d,u}$ are the proton's tensor charges.  The ratio $g_T^d/g_T^u$ is renormalisation scale invariant.   Analogous lQCD ratios in this case are: $g_T^d/g_T^u = -0.25(2)$ \cite{Bhattacharya:2016zcn, Gupta:2018lvp}; $g_T^d/g_T^u =-0.29(3)$ \cite{Alexandrou:2019brg}.
%The former value is consistent with the quark model result, whereas the latter is larger in magnitude.

The tensor charge ratio has not been calculated in the framework employed herein; but it was computed using a quark+diquark Faddeev equation built upon the sym\-metry-preserving regularisation of a contact interaction \cite{Xu:2015kta}, with the result  $g_T^d/g_T^u =-0.32(7)$, and within a three-body Faddeev equation approach \cite{Wang:2018kto}, which yielded $g_T^d/g_T^u = -0.24(1)$.  A calculation of this ratio using our framework would enable additional informative comparisons.

%The observations in this subsection bear on the proton spin puzzle, which has a thirty-year history \cite{Ashman:1987hv}.  They will subsequently be expounded upon elsewhere \cite{Chen:2020:progressSpin}.

\begin{figure}[t]
\vspace*{2ex}

\leftline{\hspace*{0.5em}{\large{\textsf{A}}}}
\vspace*{-4ex}
\includegraphics[width=0.42\textwidth]{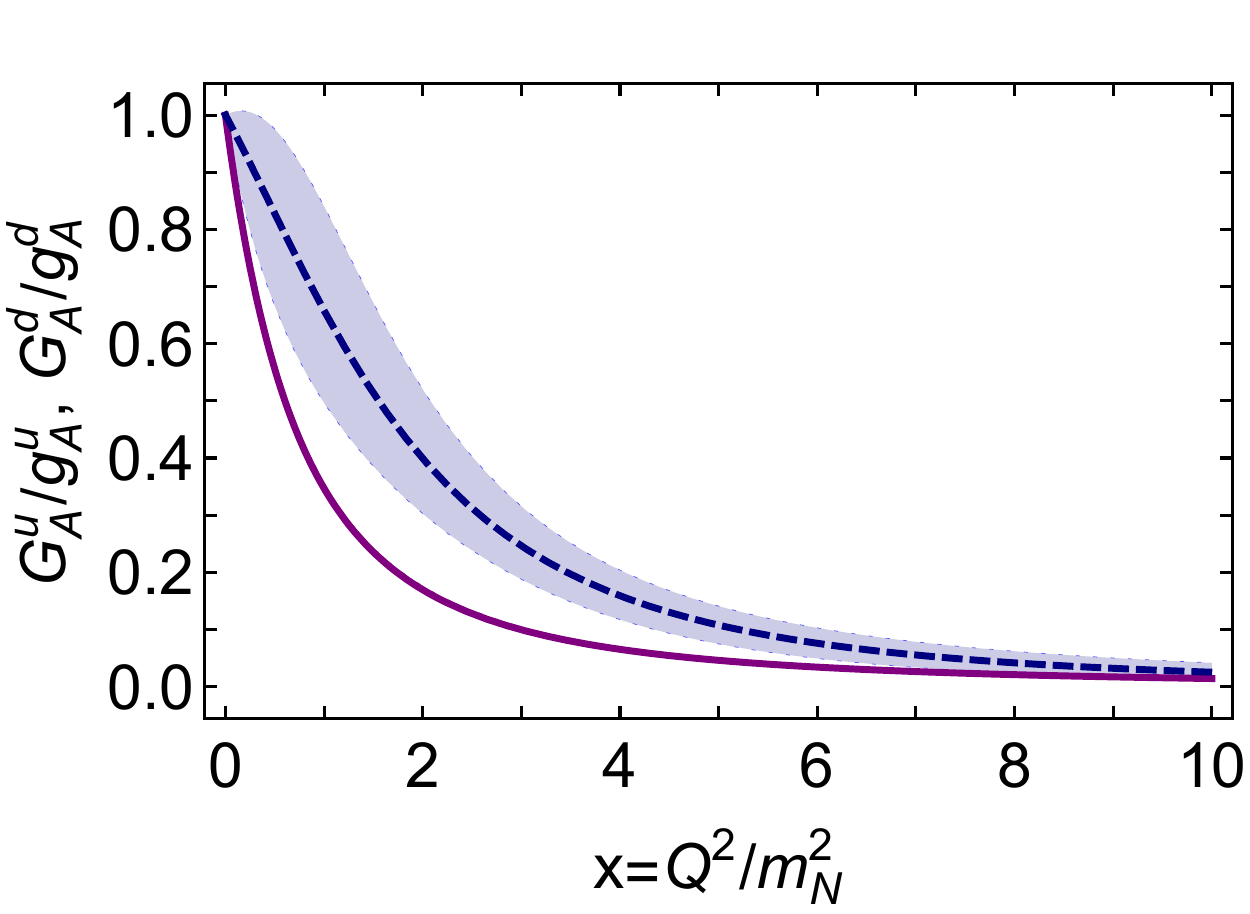}
\vspace*{0ex}

\leftline{\hspace*{0.5em}{\large{\textsf{B}}}}
\vspace*{-4ex}
\includegraphics[width=0.42\textwidth]{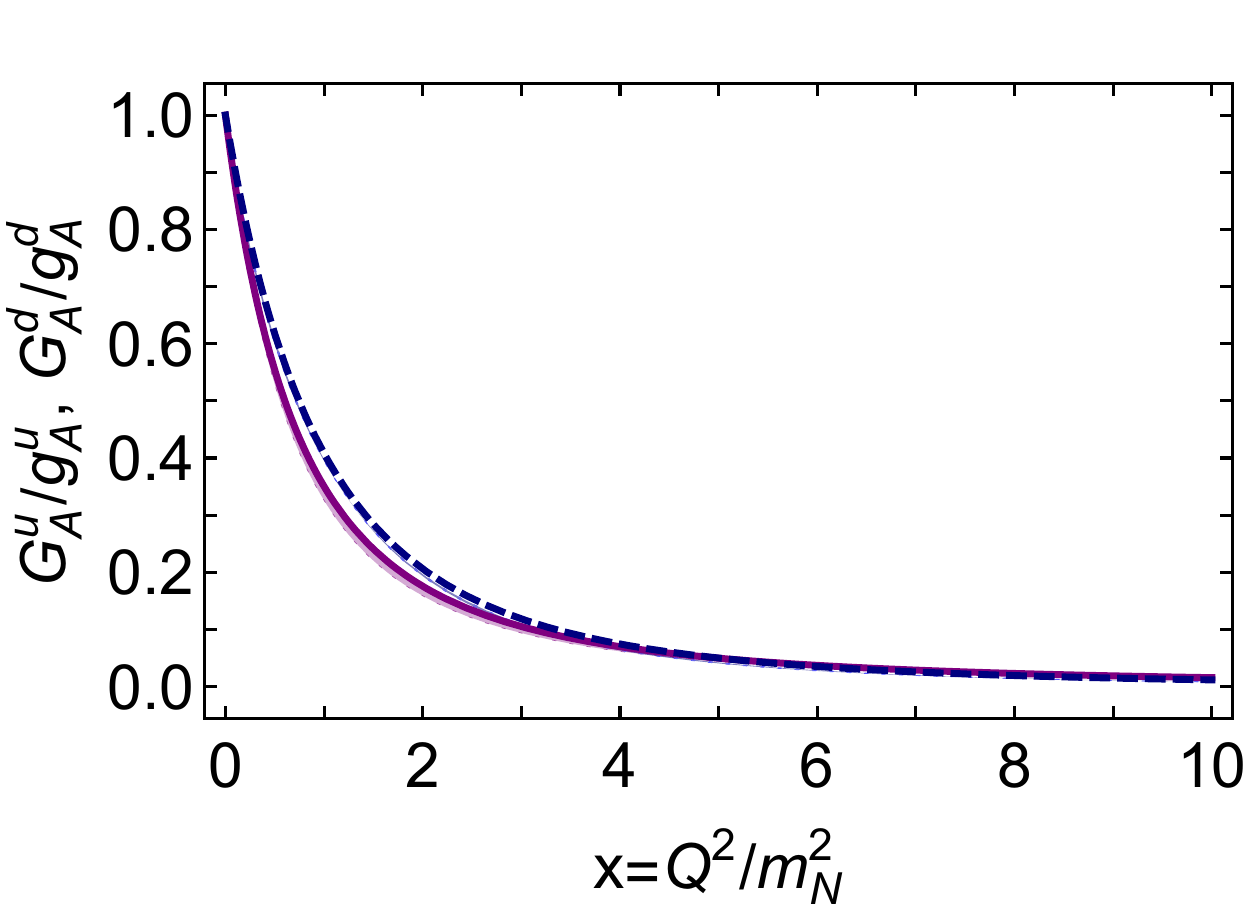}
\caption{\label{GAflavour}
Individual quark flavour contributions to proton axial form factor.
{\sf Panel A}.
Isolating the scalar-diquark part of the proton Faddeev amplitude.
Fig.\,\ref{figcurrent}\,-\,Diagram~1 -- $G_A^u(x)/g_A^u$; and Fig.\,\ref{figcurrent}\,-\,Diagram~4 -- $G_A^d(x)/g_A^d$.  In these cases, $g_A^u=0.88(7)$, $g_A^d=-0.046(7)$, $g_A^d/g_A^u = -0.054(13)$.
{\sf Panel B}.  Using complete proton Faddeev amplitude, in which case all diagrams in Fig.\,\ref{figcurrent} contribute and $g_A^u=0.94(4)$, $g_A^d=-0.30(1)$.
}
\end{figure}

Given the preceding observations, it is worth calculating and contrasting the $u$- and $d$-quark contributions to the proton axial form factor light-front transverse spatial density profiles, defined thus:
\begin{equation}
\label{density}
\hat\rho_A^f(|\hat b|) = \int\frac{d^2 \vec{q}_\perp}{(2\pi)^2}\,{\rm e}^{i \vec{q}_\perp \cdot\hat b}G_A^f(x)\,,
\end{equation}
with $G_A^f(x)$ interpreted in a frame defined by $Q^2 = m_N^2 q_\perp^2$, $m_N q_\perp = (\vec{q}_{\perp 1},\vec{q}_{\perp 2},0,0)=(Q_1,Q_2,0,0)$.  Defined this way, $|\hat b|$ and $\hat\rho_A^f $ are dimensionless.

\begin{figure*}[!t]
\hspace*{-1ex}\begin{tabular}{lcl}
\large{\textsf{A}} & & \large{\textsf{B}}\\[-3.5ex]
%
%{\sf A} &\hspace*{2em} & {\sf B} \\[-2ex]
\includegraphics[clip, width=0.4\textwidth]{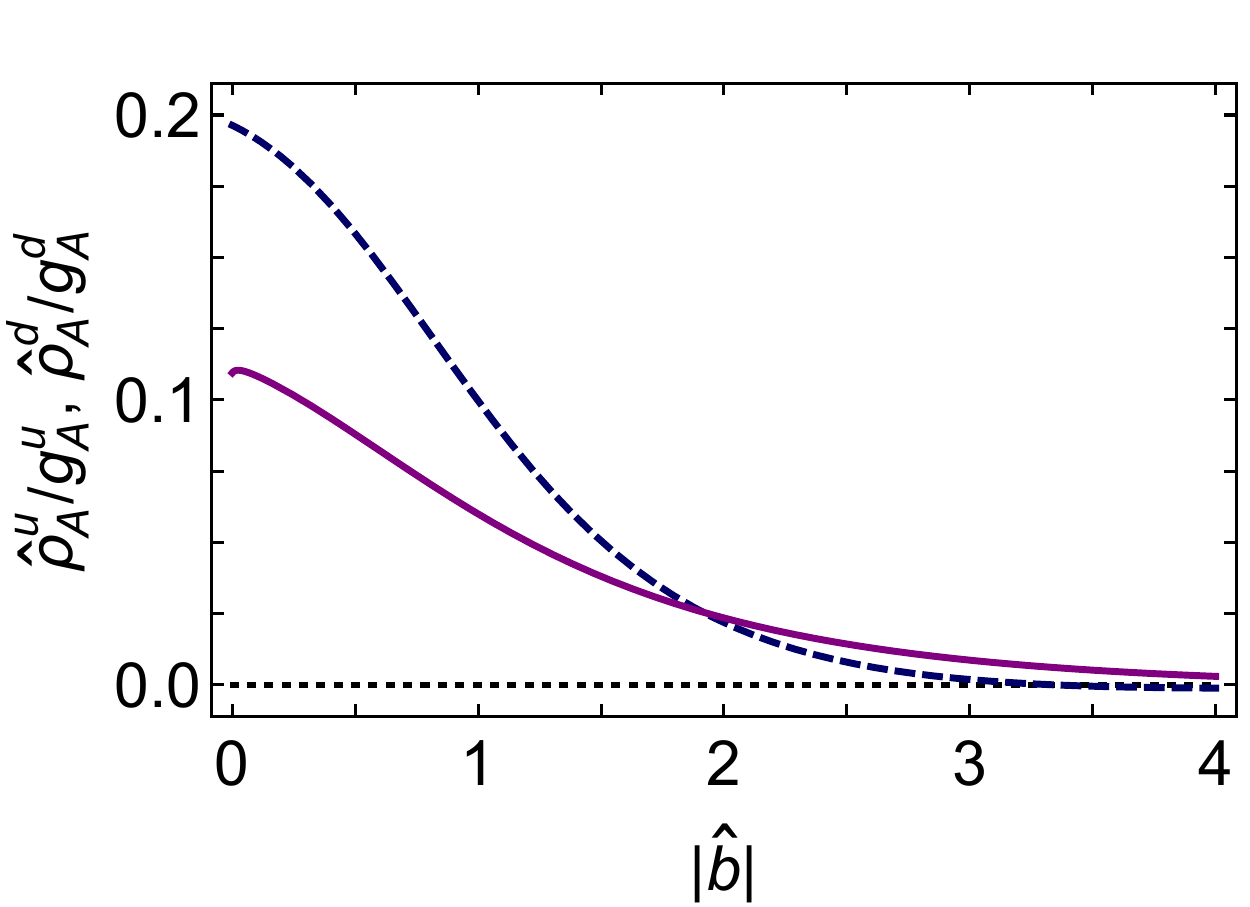} & \hspace*{4em} &
\hspace*{-0.6em}\includegraphics[clip, width=0.415\textwidth]{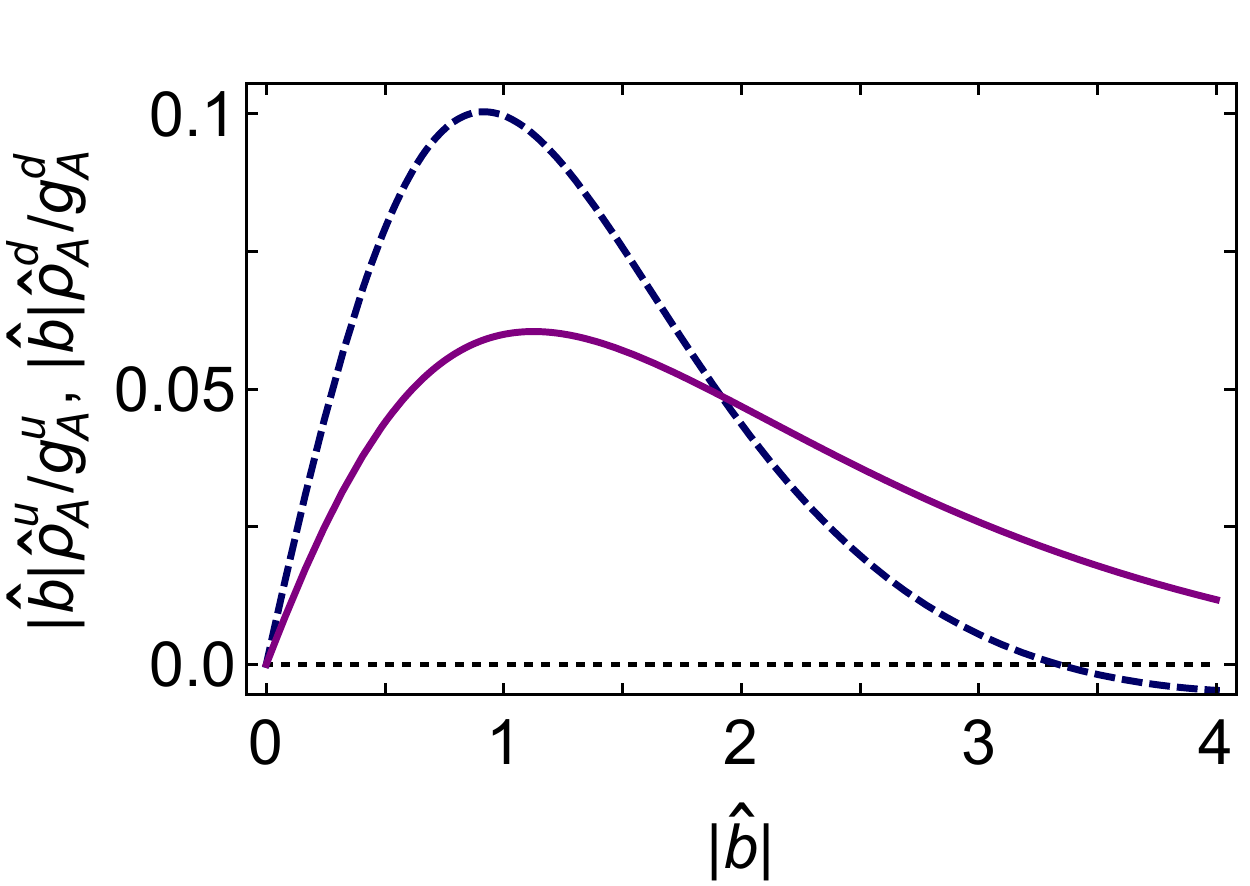} \\
\large{\textsf{C} \hspace*{1em}$0^+\!, d $} & & \large{\textsf{D} \hspace*{1em}$0^+\!, u $}\\[-4.0ex]
%
%\large{\textsf{C} \hspace*{1em}$0^+$\&$1^+\!, d $} & & \large{\textsf{D} $0^+$\&$1^+\!, u $}\\[1.0ex]
%
\includegraphics[clip, width=0.4\textwidth]{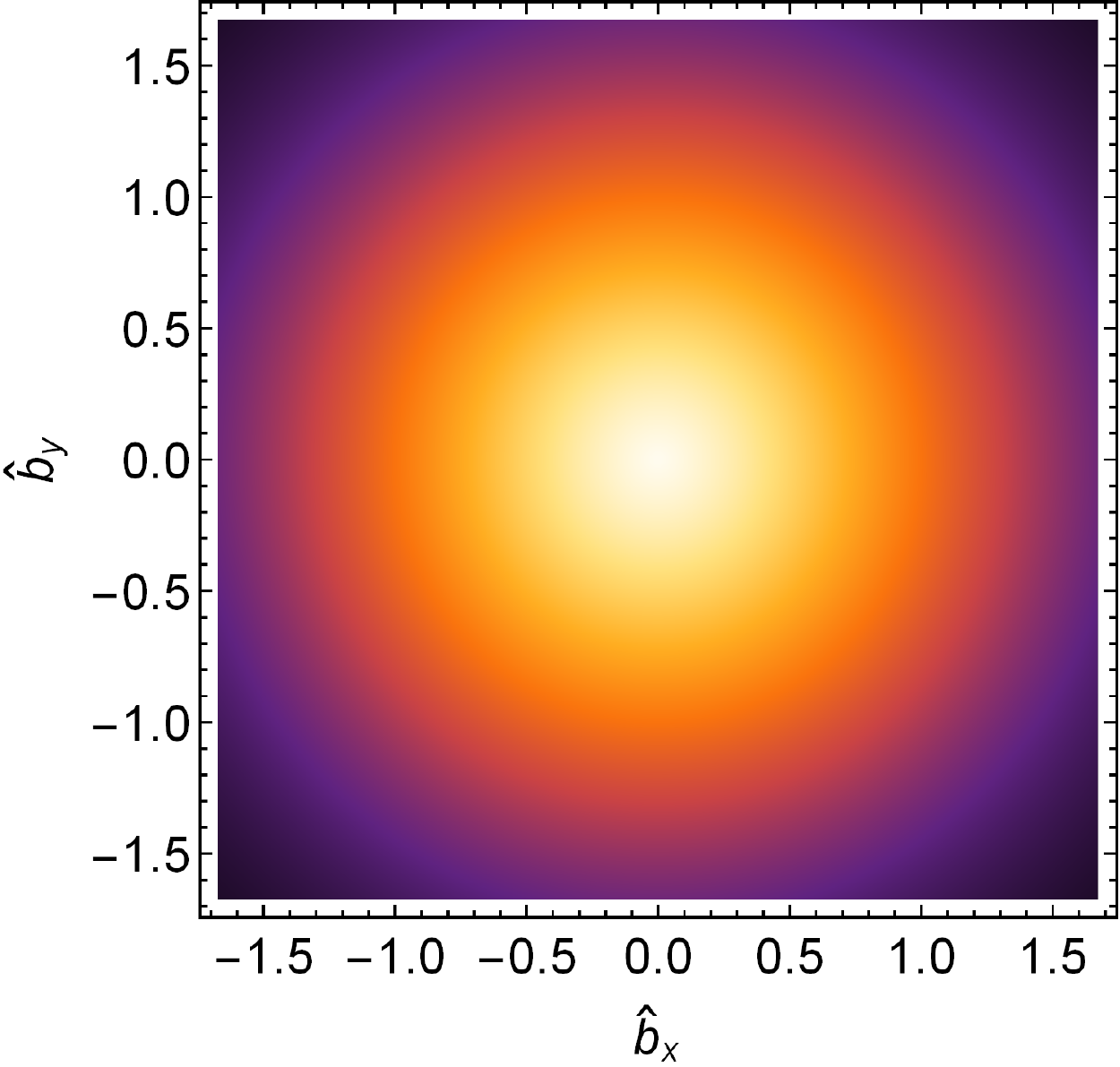} &
\hspace*{2em} \includegraphics[clip, width=0.06\textwidth]{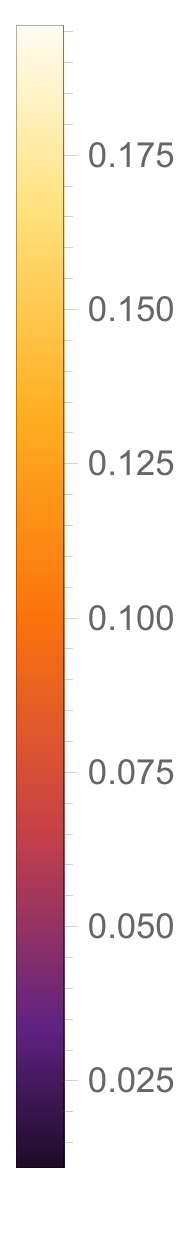} \hspace*{0em} &
\includegraphics[clip, width=0.4\textwidth]{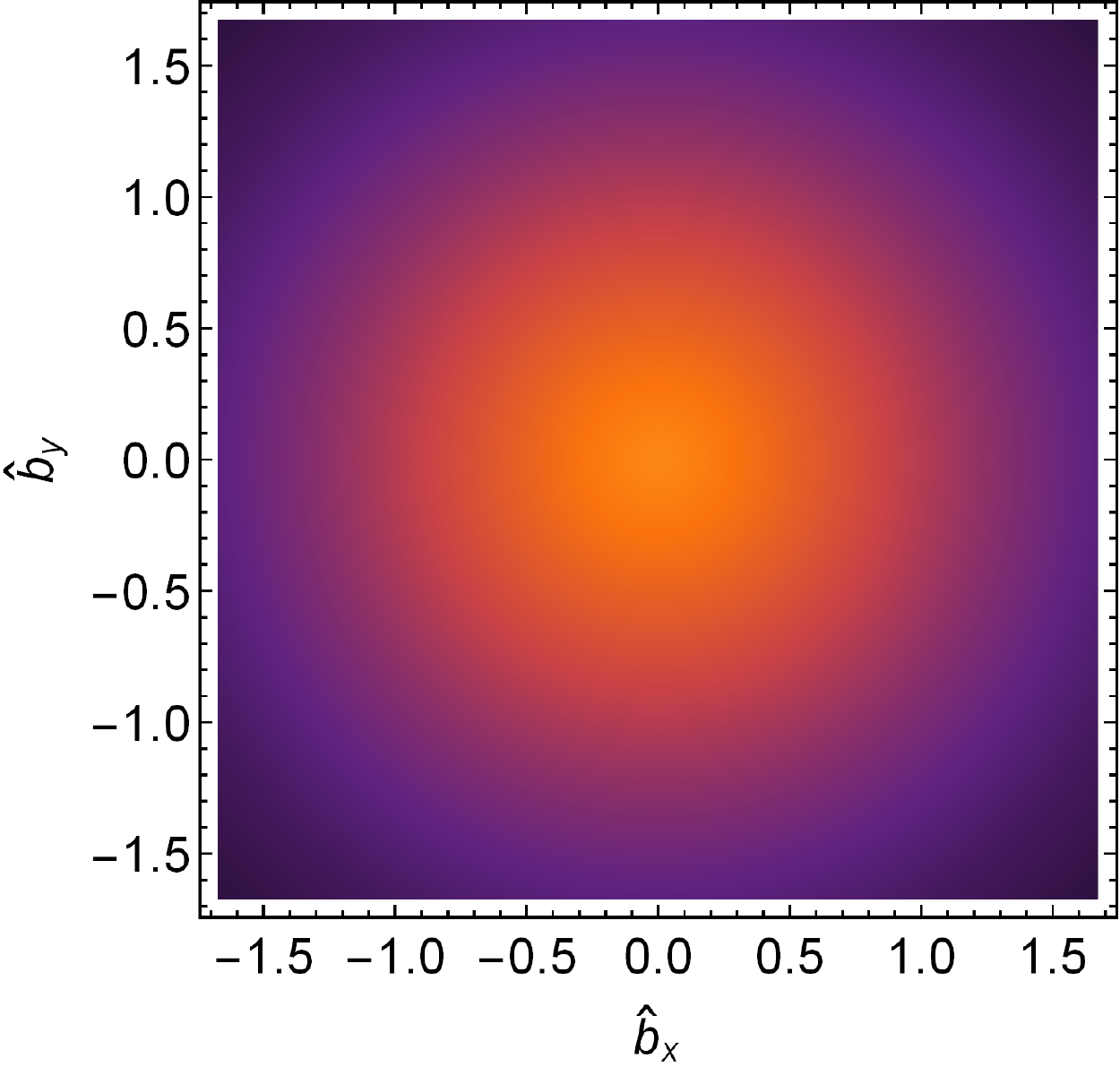}
\end{tabular}
\caption{\label{figdensity2D}
Transverse density profiles,  Eq.\,\eqref{density}, calculated from the flavour separated proton axial form factors in Fig.\,\ref{GAflavour}A, \emph{i.e}., for a scalar-diquark-only proton.
{\sf Panel A} -- $\hat\rho_A^f(|\hat b|)/g_A^f$;
{\sf Panel B} -- $|\hat b|\hat\rho_A^f(|\hat b|)/g_A^f$;
{\sf Panel C} -- two-dimensional plot of $\hat\rho_A^d(|\hat b|)/g_A^d$;
{\sf Panel D} -- similar plot of $\hat\rho_A^u(|\hat b|)/g_A^u$.
Removing the $1/g_A^f$ normalisation, the $b=0$ values are $\hat \rho_A^d(0) = -0.009$, $\rho_A^u(0)= 0.097$.
\emph{N.B}.\ $\int d^2 \hat b \,\hat\rho_A^f(|\hat b|)/g_A^f = 1$, $f=u,d$.
}
\end{figure*}

In the case where axialvector diquark correlations are omitted, one obtains the distinct flavour contributions to the proton axial form factor depicted in Fig.\,\ref{GAflavour}A.  Evidently, the $d$-quark piece, which is generated solely by Fig.\,\ref{figcurrent}\,-\,Diagram\,4, is significantly harder than the $u$-quark part, generated by Fig.\,\ref{figcurrent}\,-\,Diagram\,1.
When regarding Fig.\,\ref{GAflavour}A, the normalisation of each distribution should be borne in mind.  It was done to expose the differing profiles.  In reality, on the depicted domain, the mean value of $|G_A^d(x)/G_A^u(x)|$ is $\approx 0.1$.
%Relative to the $u$-quark contribution, the $d$-quark piece is only $5-10$\% as large in magnitude.

Reincorporating the axialvector part of the proton Faddeev amplitude, one obtains the flavour separated contributions in Fig.\,\ref{GAflavour}B.  The $d$-quark piece is now only a little harder than the $u$-quark part and the ratio of $x=0$ magnitudes is $g_A^d/g_A^u=-0.32(2)$, Eq.\,\eqref{Separations}.  On average, $|G_A^d(x)/G_A^u(x)|$ is $\approx 0.32$.
These changes are understood by noting that Fig.\,\ref{figcurrent}\,-\,Diagram\,1 now also includes spectator axialvector diquark terms, which provide contributions with similar $Q^2$-distributions for both $u$- and $d$-quarks; as do all other additional diagrams.

%\cite{CLAS:2012zia}

Working with the results depicted in Fig.\,\ref{GAflavour}A and Eq.\,\eqref{density}, one obtains the valence-quark spatial density profiles drawn in Fig.\,\ref{figdensity2D}.  We have used dimensionless quantities, which can be mapped into physical units using:
\begin{equation}
\rho_A^f(|b| =| \hat b|/m_N) = m_N^2 \,  \hat \rho_A^f(|\hat b|)\,;
\end{equation}
hence, $|\hat b| = 1$ corresponds to $|b| \approx 0.2\,$fm and $\hat\rho_1^f=0.1 \Rightarrow \rho_1^f\approx 2.3/{\rm fm}^2$.

\begin{figure*}[!t]
\hspace*{-1ex}\begin{tabular}{lcl}
\large{\textsf{A}} & & \large{\textsf{B}}\\[-3.5ex]
%
%{\sf A} &\hspace*{2em} & {\sf B} \\[-2ex]
\includegraphics[clip, width=0.4\textwidth]{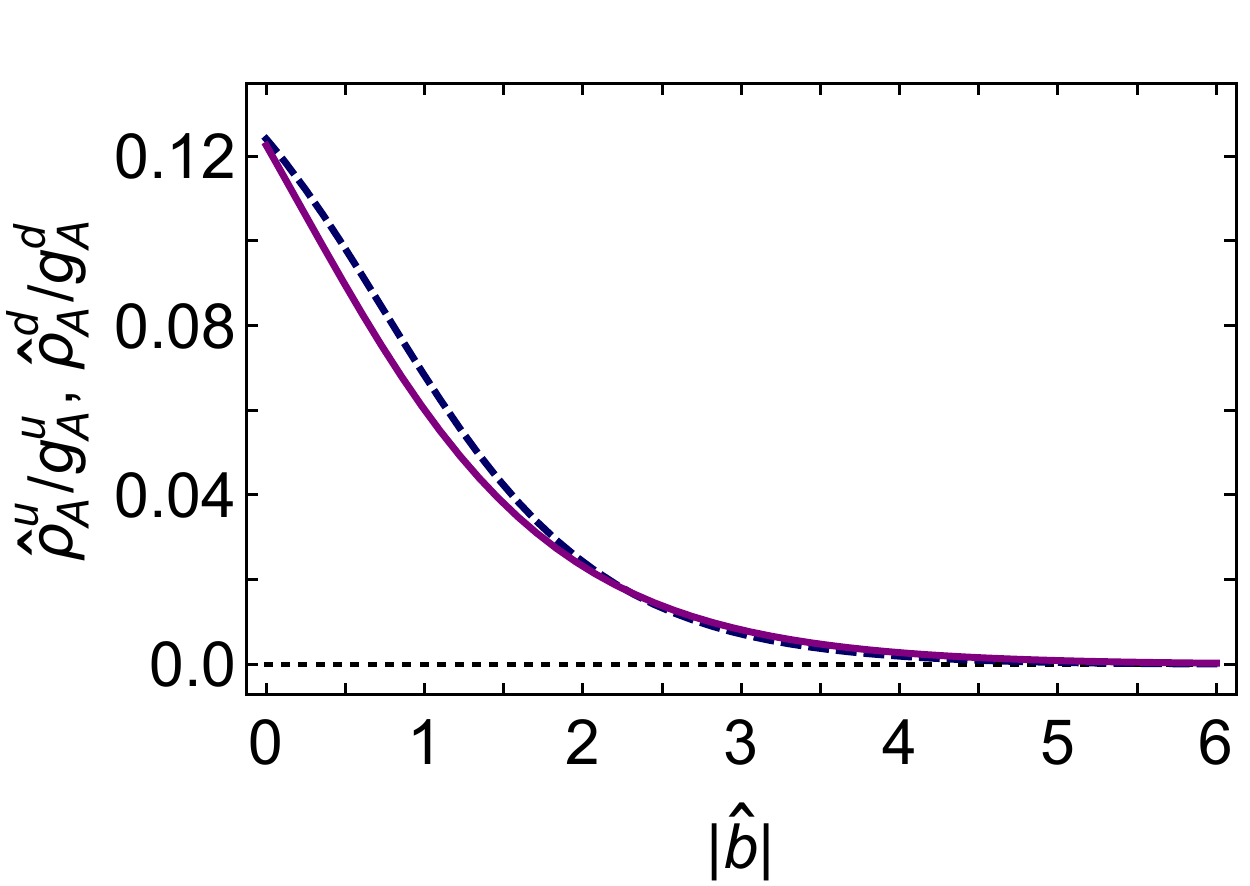} & \hspace*{4em} &
\hspace*{-0.6em}\includegraphics[clip, width=0.415\textwidth]{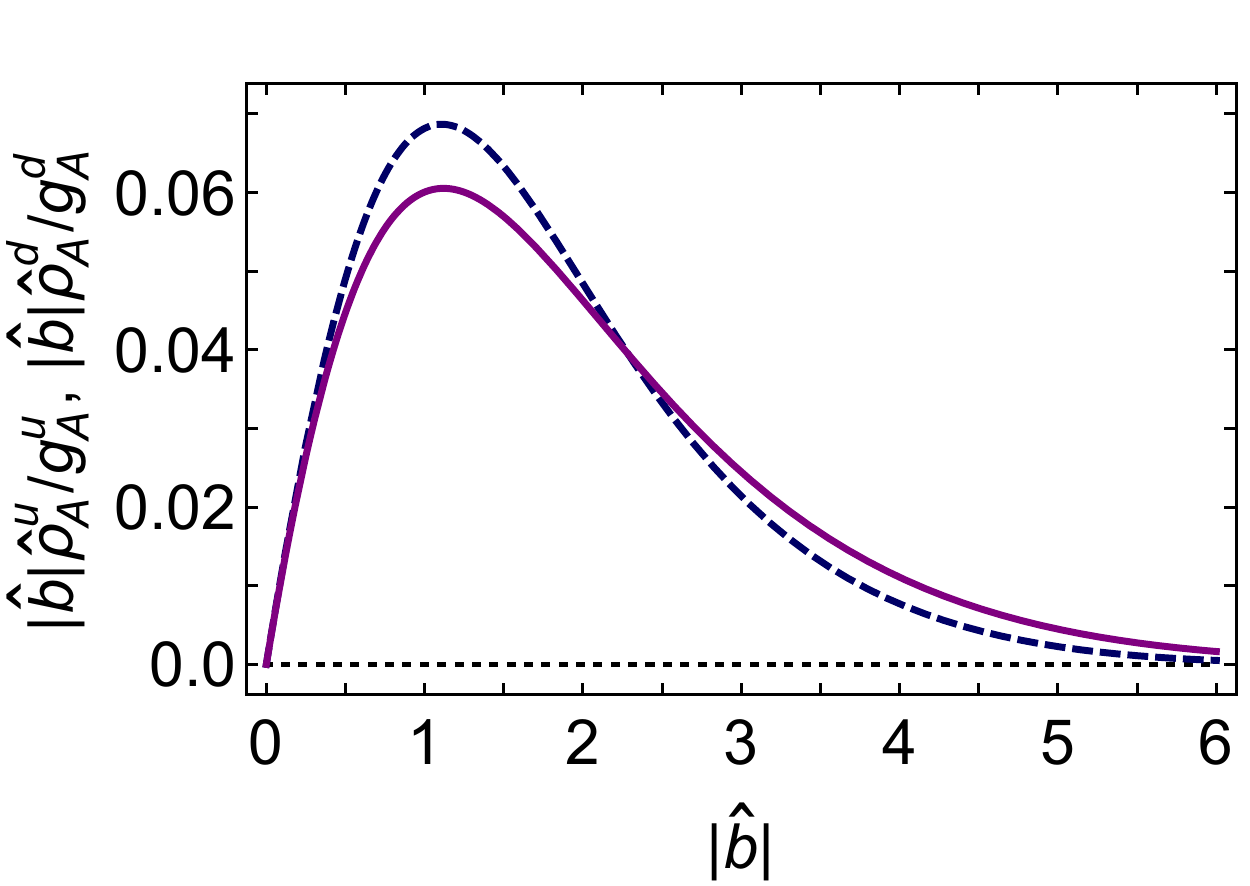} \\
%
%\large{\textsf{C} \hspace*{1em}$0^+\!, d $} & & \large{\textsf{D} \hspace*{1em}$0^+\!, u $}\\[1.0ex]
%
\large{\textsf{C} \hspace*{1em}$0^+$\&$1^+\!, d $} & & \large{\textsf{D} \hspace*{1em}$0^+$\&$1^+\!, u $}\\[-5.2ex]
\includegraphics[clip, width=0.4\textwidth]{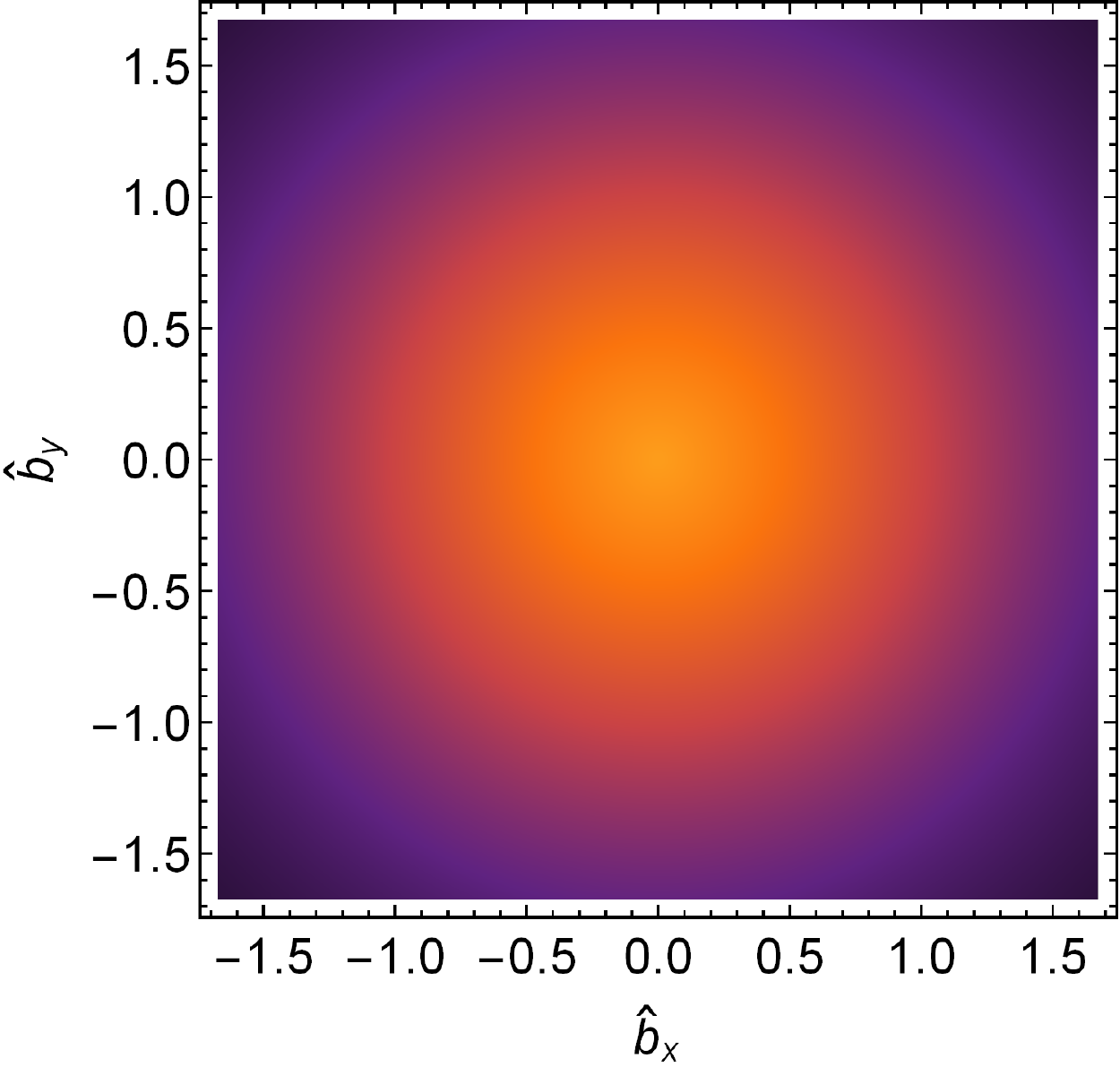} &
\hspace*{2em} \includegraphics[clip, width=0.054\textwidth]{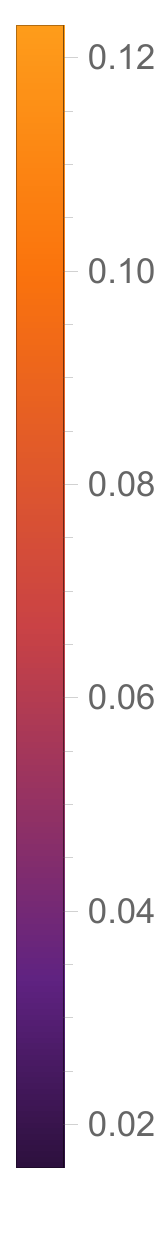} \hspace*{0.4em} &
\includegraphics[clip, width=0.4\textwidth]{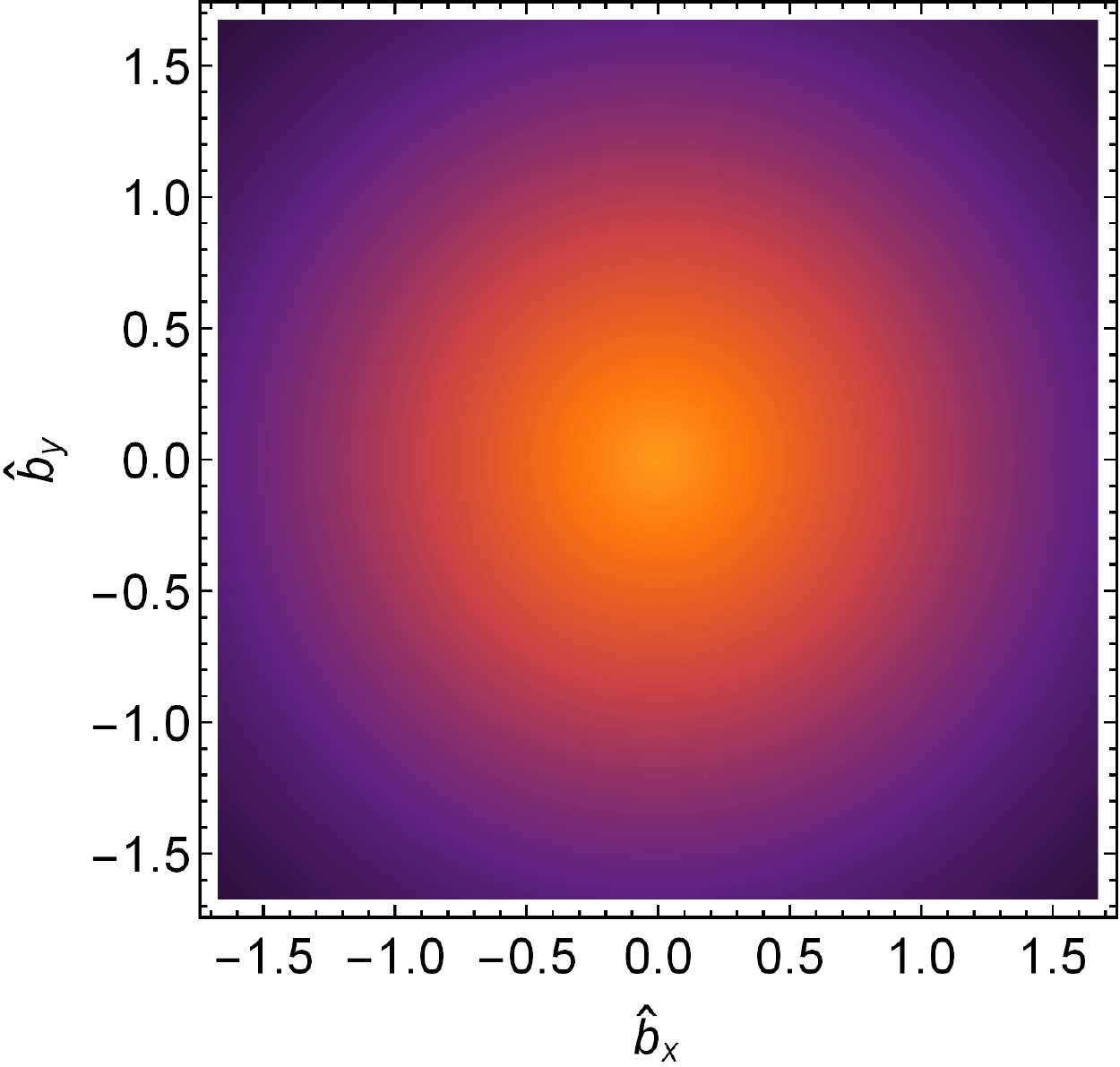}
\end{tabular}
\caption{\label{figdensity2DFull}
Transverse density profiles,  Eq.\,\eqref{density}, calculated from the flavour separated proton axial form factors in Fig.\,\ref{GAflavour}B, \emph{i.e}., for a realistic proton Faddeev amplitude.
{\sf Panel A} -- $\hat\rho_A^f(|\hat b|)/g_A^f$;
{\sf Panel B} -- $|\hat b|\hat\rho_A^f(|\hat b|)/g_A^f$;
{\sf Panel C} -- two-dimensional plot of $\hat\rho_A^d(|\hat b|)/g_A^d$;
{\sf Panel D} -- similar plot of $\hat\rho_A^u(|\hat b|)/g_A^u$.
Removing the $1/g_A^f$ normalisation, the $b=0$ values of these profiles are $\hat \rho_A^d(0) = -0.038$, $\rho_A^u(0)= 0.12$.
\emph{N.B}.\ $\int d^2 \hat b \,\hat\rho_A^f(|\hat b|)/g_A^f = 1$, $f=u,d$.
}
\end{figure*}

The top row of Fig.\,\ref{figdensity2D} depicts one-dimensional density profiles for $d$ and $u$ quarks in a scalar-diquark-only proton.
In this case, the $d$ quark profile is more pointlike than that of the $u$ quark -- $r_{A^d}^\perp = 0.24\,$fm, $r_{A^u}^\perp = 0.48\,$fm, so the $d/u$ ratio of radii is $\approx 0.5$; and $\rho_A^d(|\hat b|)$ exhibits a zero at $|\hat b| = 3.2\Rightarrow |b|=0.68\,$fm.  %, whereas the $u$ quark density is positive definite - probably
Removing the $1/g_A^f$ normalisation from these profiles, their respective $b=0$ values are $\hat \rho_A^d(0) = -0.009$, $\rho_A^u(0)= 0.097$.
The second row of Fig.\,\ref{figdensity2D} provides two-dimensional renderings of the flavour-separated transverse density profiles in the first row.  They highlight the diffuseness of the $u$ quark profile relative to that of the $d$ quark, \emph{i.e}., its greater extent in the light-front transverse spatial plane.

Turning now to Fig.\,\ref{figdensity2DFull}, the top row depicts one-dimensional density profiles for $d$ and $u$ quarks obtained when both scalar and axialvector diquarks are retained in the strength determined by the Faddeev equation in Fig.\,\ref{figFaddeev}.
In this realistic case, the $d$ quark profile is only somewhat more pointlike than that of the $u$ quark -- $r_{A^d}^\perp = 0.43\,$fm, $r_{A^u}^\perp = 0.49\,$fm, so the $d/u$ ratio of radii is $\approx 0.9$; and neither profile exhibits a zero on $|b|\lesssim 2\,$fm.  %, whereas the $u$ quark density is positive definite - probably
Removing the $1/g_A^f$ normalisation, the respective $b=0$ values of these profiles are $\hat \rho_A^d(0) = -0.038$, $\rho_A^u(0)= 0.12$.
The second row of Fig.\,\ref{figdensity2DFull} displays two-dimensional representations of the flavour-separated transverse density profiles in the first row.  They highlight that, relative to the $d$-quark profile, the intensity peak is narrower for the $u$ quark.

%\noindent\emph{9.$\;$Summary and perspectives} ---
\section{Summary and outlook}
\label{epilogue}
Beginning with a Poincar\'e-covariant quark+diquark \linebreak Faddeev equation and symmetry-preserving weak interaction current [Sec.\,\ref{SecFE}], we delivered parameter-free predictions for the nucleon axialvector form factor, $G_A(Q^2)$, on the domain $0\leq x=Q^2/m_N^2\leq 10$, where $m_N$ is the nucleon mass.  Our result agrees with all currently available data [Fig.\,\ref{FigGAx}], including that set obtained using large momentum transfer threshold pion electroproduction \cite{CLAS:2012ich}, which currently covers the range $2\lesssim x\lesssim 4$.  This experimental technique could potentially be used to reach higher $Q^2$ values.

One other calculation of $G_A(Q^2)$ at large-$Q^2$ is available \cite{Anikin:2016teg}, with results on $1 \lesssim Q^2/m_N^2 \lesssim 10$.  Having used light-cone sum rules (LCSR), low-$Q^2$ was inaccessible.  Compared with our predictions, the LCSR results are pointwise different and, on average, approximately 40\% smaller [Fig.\,\ref{LCSRdiff}A].

Regarding the oft-used dipole \emph{Ansatz}, we showed that it could be used to provide a reasonable representation of $G_A(x)$ on $x\in[0,3]$.  Outside the fitted domain, however, the quality of approximation deteriorates quickly, with the dipole overestimating the true result by 56\% at $x=10$ [Fig.\,\ref{LCSRdiff}B].

We discussed the separation of the proton $G_A$ into separate contributions from valence $u$ and $d$ quarks, relating the results to the zeroth moments of the associated light-front helicity distributions; and exploiting the $Q^2$ coverage of our predictions, we also calculated the flavour-separated light-front transverse spatial density profiles [Sec.\,\ref{SecFlavourSep}].

%Given predictions on such a large $Q^2$ domain, we were able to provide a detailed analysis of the flavour separation of the proton $G_A$ into contributions from valence $u$ and $d$ quarks, relating this to the zeroth moments of the associated light-front helicity distributions, and also the flavour-separated light-front transverse spatial density profiles [Sec.\,\ref{SecFlavourSep}].

Our value of $g_A^d/g_A^u=-0.32(2)$ is consistent with available experimental data, but significantly lower in magnitude than recent results from lattice-regula\-rised QCD.
On the other hand, comparing magnitudes, it is markedly larger than the value typical of nonrelativistic constituent quark models with uncorrelated wave functions ($-1/4$).  The enhancement owes to the presence of axialvector diquark correlations in our Poincar\'e-covariant nucleon wave function.  Importantly, in the absence of axialvector diquarks, $g_A^d/g_A^u=-0.054(13)$.

Our calculated light-front transverse density profiles revealed that, omitting axialvector diquarks, the magnitude of the $d$ quark contribution to $G_A$ is just 10\% of that from the $u$ quark and the $d$ quark is also much more localised [Fig.\,\ref{figdensity2D}].
Working instead with a realistic axialvector diquark fraction, the $d$ and $u$ quark transverse profiles are quite similar, after accounting for their different normalisations [Fig.\,\ref{figdensity2DFull}].

Owing to the importance of developing a theoretical understanding of nucleon spin structure, it would be worth extending the work on $g_A^d$, $g_A^u$ described herein to the analogous problem of proton tensor charges, whose precision measurement is a high-profile goal \cite{Ye:2016prn}.  Equally and vice versa, a three-quark Faddeev equation treatment of the proton's flavour-separated axial charges would be valuable, following Ref.\,\cite{Wang:2018kto} and possibly improving upon that study by using a more sophisticated bound-state kernel of the type discussed elsewhere \cite{Qin:2020jig}.

More immediately, one can readily adapt the Faddeev equation and current used in the present study to weak interaction induced transitions between the nucleon ($N$) and its lowest lying excitations, such as the $\Delta$-baryon and the $N^\ast(1535)$.  Many calculations of the $N\to \Delta$ transition exist, using a variety of frameworks.  Moreover, this process is important for a reliable understanding of neutrino scattering.  Hence, completing a fully Poincar\'e-covariant analysis of the weak $N\to \Delta$ transition, on a domain that stretches from low-to-large $Q^2$, is a high priority.
%  Aliev:2019tmk parity partner.

%
%\section*{Acknowledgments}
\medskip
\noindent\emph{Acknowledgments}.
We are grateful for constructive comments from Z.-F.~Cui, C.\,S.~Fischer, V.\,I.~Mokeev, J.~Se\-go\-via and B.~Wojtsekhowski.
Work supported by:
National Natural Science Foundation of China (grant nos.\ 12135007 and 12047502).
%
%and
%
%Natural Science Foundation of Jiangsu Province (grant no.\ BK20220323).

%\bibliographystyle{../../../zProc/z10/z10KITPC/h-physrev4}
%%\bibliographystyle{elsarticle-num-names}
%\bibliography{../../../../CollectedBiB}

\end{document}